\documentclass[12pt,preprint,psfig,epsf]{aastex}

\newcommand{\dmm}{\mbox{$\Delta$m$_{15}(B)$}}
\newcommand{\ubvri}{\protect\hbox{$U\!BV\!RI$} }
\newcommand{\bvri}{\protect\hbox{$BV\!RI$} }
\newcommand{\bvrij}{\protect\hbox{$BV\!RIJ$} }
\newcommand{\bvrijhk}{\protect\hbox{$BV\!RIJHK$} }
\newcommand{\yjhk}{\protect\hbox{$Y\!JHK$} }

\shorttitle{The Type~Ia Supernova 2004S}
\shortauthors{Krisciunas et al.} 

\begin{document}
\received{12 May 2006}

\title{The Type Ia supernova 2004S, a clone of SN~2001el,
and the optimal photometric bands for extinction estimation\altaffilmark{1}}

\author{
Kevin Krisciunas,\altaffilmark{2}
Peter M. Garnavich,\altaffilmark{2}
Vallery Stanishev,\altaffilmark{3}
Nicholas B. Suntzeff,\altaffilmark{4}
Jose Luis Prieto,\altaffilmark{5}
Juan Espinoza,\altaffilmark{6} 
David Gonzalez,\altaffilmark{6} 
Maria Elena Salvo,\altaffilmark{7}
Nancy Elias de la Rosa,\altaffilmark{8,9}
Stephen J. Smartt,\altaffilmark{10}
Justyn R. Maund,\altaffilmark{11} and
Rolf-Peter Kudritzki\altaffilmark{12}
}
\altaffiltext{1}{Based in part on observations taken at the Cerro Tololo
Inter-American Observatory, National Optical Astronomy Observatory, 
which is operated by the Association of Universities for Research in 
Astronomy, Inc. (AURA) under cooperative agreement with the National 
Science Foundation.}
\altaffiltext{2}{Department of Physics, University of Notre Dame, 225
  Nieuwland Science Hall, Notre Dame, IN 46556-5670;
  {kkrisciu@nd.edu}, {pgarnavi@nd.edu} }
\altaffiltext{3}{Department of Physics, Stockholm University, AlbaNova
 University Center, SE-106 91 Stockholm, Sweden; {vall@physto.se} }
\altaffiltext{4}{Department of Physics, Texas A. \& M. University,
  College Station, TX 77843; {suntzeff@physics.tamu.edu} }
\altaffiltext{5}{Department of Astronomy, Ohio State University,
  4055 McPherson Laboratory, 140 W. 18th Ave., Columbus, Ohio 43210;
   {prieto@astronomy.ohio-state.edu} }
\altaffiltext{6}{Cerro Tololo Inter-American Observatory, Casilla 603,
  La Serena, Chile; {jespinoza@ctio.noao.edu}, {dgonzalez@ctio.noao.edu} }
\altaffiltext{7}{The Research School of Astronomy and Astrophysics,
  The Australian National University, Mount Stromlo and Siding Spring
  Observatories, via Cotter Rd, Weston Creek PO 2611, Australia;
  {salvo@mso.anu.edu.au} }
\altaffiltext{8}{INAF-Osservatorio Astronomico de Padova, Vicolo dell'Osservatorio 5, 
  I-35122 Padova, Italy; {nancy.elias@oapd.inaf.it} }
\altaffiltext{9}{Universidad de La Laguna, Ave. Astrof\'{i}sico Francisco S\'{a}nchez s/n, 
E-38206. La Laguna, Tenerife, Spain}
\altaffiltext{10}{Department of Physics and Astronomy, Queen's University
  Belfast, Belfast BT7 1NN, Northern Ireland, UK; {s.smartt@qub.ac.uk} }
\altaffiltext{11}{University of Texas, McDonald Observatory, 1 University Station 
  C1402, Austin, TX 78712-0259 ; {jrm@astro.as.utexas.edu} }
\altaffiltext{12}{Institute for Astronomy, University of Hawaii, 2680
  Woodlawn Drive, Honolulu, HI 96822; {kud@ifa.hawaii.edu} }

\begin{abstract} 

%[Version of 18 August 2006]
We present optical (\ubvri\hspace{-1.0mm}) and near-infrared
(\yjhk\hspace{-1.0mm})  photometry of the normal Type Ia supernova 2004S.  We
also present eight optical spectra and one near-IR spectrum of SN 2004S.  The light
curves and spectra are nearly identical to those of SN~2001el.  This is the
first time we have seen optical {\em and} IR light curves of two Type Ia
supernovae match so closely. Within the one parameter family of light curves
for normal Type Ia supernovae, that two objects should have such similar light
curves implies that they had identical intrinsic colors and produced similar
amounts of $^{56}$Ni. From the similarities of the light curve shapes we obtain
a set of extinctions as a function of wavelength which allows a simultaneous
solution for the distance modulus difference of the two objects, the difference
of the host galaxy extinctions, and R$_V$. Since SN~2001el had roughly an order
of magnitude more host galaxy extinction than SN~2004S, the value of R$_V$ =
2.15$^{+0.24}_{-0.22}$ pertains primarily to dust in the host galaxy of
SN~2001el.  We have also shown via Monte Carlo simulations that adding rest
frame $J$-band photometry to the complement of \bvri photometry of Type Ia SNe
decreases the uncertainty in the distance modulus by a factor of 2.7.  A
combination of rest frame optical and near-IR photometry clearly gives more
accurate distances than using rest frame optical photometry alone.
\end{abstract}


\section{Introduction}

A Type Ia supernova (SN, plural SNe) is widely believed to be a carbon-oxygen
white dwarf which explodes when its mass reaches the Chandrasekhar limit owing to
mass transfer from a nearby companion \citep[][and references therein]{Liv00}.  
The explosion produces several tenths of a solar mass of radioactive $^{56}$Ni,
which decays to $^{56}$Co, and finally to stable iron \citep{Str_etal06}.  Since
the progenitors of Type Ia SNe have approximately the same mass
(M$_{Ch} \approx$ 1.4 M$_{\odot}$), the explosions
have approximately the same brightness at maximum light.  \citet{Phi93} first
demonstrated unambiguously that the peak brightness of a Type Ia SN is related to
the decline rate of the light curve.  Since the publication of Phillips' classic
paper, the common measure of the decline rate has been the number of $B$-band
magnitudes that a Type Ia SN fades in the first 15 days after maximum light
(designated \dmm).  A typical value is \dmm = 1.10 mag.  Values range from
$\sim$0.7 for SN~2001ay (Nugent et al. 2006, in preparation) to 1.93 for
SN~1991bg \citep{Fil_etal92,Lei_etal93,Phi_etal99}.

In the past 13 years supernova researchers have noted various patterns in the
light curves of Type Ia SNe.  This has allowed the determination of the peak
magnitudes, host galaxy extinction, and distances to these objects even if the
data of a particular object are irregularly spaced in time and do not necessarily
overlap the light curve maxima.  The three most widely used systems for fitting
optical light curves are the \dmm\ system itself \citep{Ham_etal96, Phi_etal99,
Ger_etal04, Pri_etal06}, the ``stretch method'' \citep{Per_etal97, Gol_etal01,
Kno_etal03}, and the multi-color light curve shape (MLCS) method
\citep{Rie_etal96a, Rie_etal98, Jha_etal06b}.  There is also the hybrid 
magnitude-color system of \citet{Wan_etal03a}.

Host galaxy extinction of Type Ia SNe is measured empirically by
first determining reddening-free color loci, determining color
excesses, and using empirically determined scale factors to convert
color excesses to A$_V$ or extinction in some other band.  \citet{Lir95}
found, from a small sample of Type Ia SNe unreddened in their hosts,
that $B-V$ colors are uniform and evolve in the same way from 30
to 90 days after the time of $V$-band maximum.  \citet{Nob_etal03}
investigated the optical colors of Type Ia SNe and found that all the
colors except $V-I$ are consistent with zero intrinsic dispersion at
35 days after T($B_{max}$).  \citet{Wan_etal05} found that the $B-V$
colors at 12 days after T($B_{max}$) provided the best method of determining
the $B-V$ color excess.  With MLCS2k2 \citet{Jha_etal06b} adopt an intrinsic
color of $B-V$ = 1.054 $\pm$ 0.049 at $t$ = 35 days and assume an exponential
distribution of host galaxy reddening with scale length E($B-V$) = 0.138 mag.
They also adopt a Bayesian prior on the distribution of R$_V$, with a minimum value
of 1.8.  Then they marginalize over the optical photometry to obtain the
best values of A$_V$ and R$_V$.

The prime problem with using optical photometry to determine A$_V$ along the line
of sight to a Type Ia SN can be illustrated as follows.  For normal dust in our
Galaxy R$_V$ $\equiv$ A$_V$ / E($B-V$) $\approx$ 3.1 \citep{Car_etal89}.  A small
uncertainty in the $B-V$ color excess translates into a much larger uncertainty
in A$_V$.  The scale factor R$_V$ is not necessarily equal to 3.1.  The mean
value most appropriate to galaxies that host Type Ia SNe is somewhere in the
range 2.3 to 2.65 \citep{Rie_etal96b, Alt_etal04, Rei_etal05, Wan_etal06,
Jha_etal06b}. Some very low values have been measured.  The Type Ia SN 1999cl was
reddened by dust with R$_V$ = 1.55 $\pm$ 0.08 \citep {Kri_etal06}.  The dust in
the host of SN~2003cg had R$_V$ = 1.80 $\pm$ 0.19 \citep{Eli_etal06}.  Our
observations of SN~2006X are consistent with R$_V$ $\approx$ 
1.56.\footnote[13]{See http://www.nd.edu/~kkrisciu/sn2006X.html.}
Perhaps these very low values of R$_V$ associated with highly reddened objects
are the result of an interaction of the SN ejecta and the surrounding circumstellar
material \citep{Wan05}.

To account for non-standard reddening laws and to obtain the most robust
value of the extinction towards a Type Ia SN, it is best to combine optical
and IR photometry.  \citet{Eli_etal85} first suggested that the $V-K$ colors
of Type Ia SNe might be uniform and therefore very useful for determining
extinction for these objects. Using the standard reddening law of
\citet{Car_etal89}, A$_V$ $\approx c_R$ E($V-K$), where $c_R$ = 1.13 for
Galactic (R$_V$ = 3.1) dust, and $c_R$ = 1.07 for the extreme case of
R$_V$ = 1.55 for SN~1999cl.  Thus, if we can determine a
$V-K$ color excess, it needs to be scaled only slightly to give us A$_V$.
Using data of 8 Type Ia SNe, we determined zero-reddening $V-J$, $V-H$, and $V-K$
loci of Type Ia SNe with mid-range decline rates \citep{Kri_etal00}.  
Subsequently, we determined that the corresponding loci of Type Ia SNe with
slow decline rates (\dmm\ $\lesssim$ 1.0) are roughly 0.24 mag bluer
\citep{Kri_etal04b}. We note that $V-J$ colors of Type Ia SNe are the least
uniform of the three color indices.  Also, our previous loci are valid for
the time range $-$9 $\leq t \leq$ +27 days, where $t$ is measured with
respect to the time of $B$-band maximum.

Since the Cal\'{a}n-Tololo SN survey in the early 1990's, extensive efforts
have been devoted to discovering and following the light curves of Type Ia
SNe, most notably the SN search carried out with the Katzman Automatic
Imaging Telescope at Lick Observatory \citep{Fil_etal01}. Because of
the time required to observe the SNe, obtain image subtraction templates for
those objects that require templates, and to reduce the data for many
objects, one recent trend has been to publish papers on individual objects
which are unusual, such as SNe~2000cx \citep{Li_etal01,Can_etal03} and
2002cx \citep{Li_etal03}.  The analogy is that pathologists can better
understand how normal human organs work by studying organ malfunction and
disease.  In this paper we discuss the astronomical case of SN~2004S, a
normal Type Ia SN, the light curves of which are essentially clones of another
object, SN~2001el \citep{Kri_etal03}.

SN~2004S was discovered using images taken on 2004 February 3.54 and 4.56 UT by
\citet{Mar04}.  The supernova was located at RA = 06$^h$ 45$^m$
43.\hspace{-1.3mm}$^s$5, DEC =$-$31$^{\rm o}$ 13$^{\prime}$ 52$\farcs$5
(J2000), some 47$\farcs$2 west and 2$\farcs$5 south of the core of the Sc
galaxy MCG -05-16-021 \citep{Big04}. SN~2004S was confirmed to be a Type~Ia
supernova by \citet{Sun_etal04} from a spectrum taken on 2004 Feb 6.1 UT with
the CTIO 1.5-m telescope.  In Fig. \ref{comp_at_max} we see the
\citet{Sun_etal04} spectrum, along with a spectrum of SN~2001el obtained by
\citet{Wan_etal03b}. Given the similarities of the optical and IR light curves
and the spectra of these two objects, we feel fully justified in calling these
two objects clones.

According to \citet{The_etal05}, the heliocentric radial velocity of
MCG -05-16-021 is 2808 km s$^{-1}$. In the reference frame of the Cosmic
Microwave Background (CMB) radiation the recession velocity is 2957 km
s$^{-1}$.  Assuming a Hubble constant of 72 km s$^{-1}$ Mpc$^{-1}$
\citep{Fre_etal01}, the distance modulus of SN~2004S and its host is $m-M$ =
33.07 $\pm$ 0.21 mag, where the uncertainty corresponds to an assumed random
velocity of 300 km s$^{-1}$.

The Galactic reddening along the line of sight to SN~2004S is E($B-V$) =
0.101 $\pm$ 0.010 mag \citep{Sch_etal98}. Assuming R$_V$ = 3.1 for Galactic dust, 
there must be at least 0.313 $\pm$ 0.031 mag of $V$-band extinction toward 
SN~2004S.  Given the offset of the SN from the core of
its host (see Fig. \ref{finder}), we expected the SN to be essentially 
unreddened by the host.  An
analysis of SN~2004S may thus provide useful information on the
intrinsic colors of normal Type Ia SNe of mid-range decline rates.

In this paper we present optical and near-IR photometry of SN~2004S. Other
optical photometry of the object has been published by \citet{Mis_etal05}.  
These authors obtained a light curve solution using their own measurements
plus our optical data from the CTIO 1.3-m published here.  Important conclusions 
can be made about SN~2004S by using IR data as well.  That the optical and IR light
curves of SNe 2001el and 2004S are so similar sheds light on the uniformity
of explosion models of normal Type Ia SNe and also allows us to investigate
the interstellar dust content of the hosts of SNe 2001el and 2004S beyond
what is possible from a study of either object on its own.

The paper is organized as follows.  In \S2 we present optical and infrared
photometry of SN~2004S.  In \S3 we present optical and IR spectra of SN~2004S
and compare them to several other Type Ia SNe.  In \S4 we consider the
extinction suffered by SNe 2004S and 2001el as two separate problems
and also as a joint exercise.  Using information about the relative
uncertainties of the filter by filter photometry from \S4, in \S5 we 
carry out Monte Carlo simulations of extinction and reddening for a
general model of the effect of the dust on the uncertainty of distance
measurements of Type Ia SNe.  This is relevant for future ground-based
and space-based SN surveys.  Our conclusions are presented in \S6.

\section{Optical and Infrared Photometry}

Most of our photometry was taken with the CTIO 1.3-m telescope and the
optical/IR imager ANDICAM. ANDICAM contains standard Johnson $UBV$ filters,
Kron-Cousins $R$ and $I$ filters and standard Caltech/CTIO $JHK$ filters.  
Read out in 2 $\times$ 2 binning mode, ANDICAM gives a plate scale on the
1.3-m telescope of 0$\farcs$369 px$^{-1}$ for 
optical imaging and 0$\farcs$274 px$^{-1}$ for IR imaging.
The optical field of view was 6$\farcm$3 by 6$\farcm$3, 
while the IR field of view was 2$\farcm$34 by 2$\farcm$34.

ANDICAM also contains a 1.03 $\mu$m filter (known as $Y$).  This new
filter exploits a previously unused atmospheric window.  For more
information see \citet{Hil_etal03}.  However, there are problems with
their calibration of standard stars.  From synthetic photometry of
Kurucz model spectra spanning a range of temperatures \citet[][Appendix
C]{Ham_etal06} obtained a relationship between $Y-K_s$ colors and the
published $J-K_s$ colors of \citet{Per_etal98} standards. The resulting
$Y$-band magnitudes for standard stars are typically accurate to better
than 0.01 mag.

%Previously, we have found it more reliable to obtain $Y$-band
%magnitudes of standards using the following relationship: $(Y - K_s) = -0.013 +
%1.614 (J_s - K_s)$ \citep{Kri_etal04b}.  This relationship is based on synthetic
%photometry of Vega, Sirius, and the Sun and uses the $J_s$ and $K_s$ magnitudes
%of \citet{Per_etal98}.

This paper also contains some optical photometry obtained with the Siding
Spring 2.3-m telescope and the ESO 2.2-m telescope.  One night of $JHK$
photometry was obtained with the 3.58-m Telescopio Nazionale Galileo (TNG)
at La Palma.  Two epochs of late-time near-IR photometry were obtained with
the number 1 Very Large Telescope (VLT) at Cerro Paranal, using ISAAC.

In Fig. \ref{finder} we show an optical finding chart of the field of
SN~2004S.  We calibrated the \ubvri magnitudes of some of the fields stars
on 12 photometric nights using \citet{Lan92} standards, in particular the
Rubin 149 field.  The optical field star magnitudes are to be found in Table
\ref{standards}. The near-IR magnitudes of some of these stars are to be
found in Table \ref{ir_stds}.  The IR calibration was carried out on 8
photometric nights with good seeing using the star P9123 from the list of
\citet{Per_etal98} standards.  For P9123 we estimate $Y$ = 11.077 $\pm$
0.010 from the fifth order polynomial of \citet{Ham_etal06}.
Our $JHK$ photometry of the field stars is in very good agreement (often $\pm$ 0.01 
mag) with values from the Two Micron All-Sky Survey (2MASS).

We present \ubvri photometry of SN~2004S in Table \ref{opt_photom} and
the \yjhk photometry in Table \ref{ir_photom}. The brightness and
location of the SN did not require the use of host galaxy subtraction
templates.  Our CTIO 1.3-m photometry is based on aperture photometry
using a typical aperture of radius 10 px.  On nights of bad seeing a
larger software aperture was used.

The 12 dates marked with an asterisk in Table
\ref{opt_photom} were photometric nights at CTIO.  On those occasions we
tied the SN~2004S photometry directly to the Landolt system using the Ru
149 field.  Given that the SN~2004S field and Ru 149 were always observed
at an airmass difference of less than 0.2, and given that the range of
color of the Ru 149 standards was reasonably large ($-0.129 \leq B-V \leq
1.115$) we feel that the calibration of the field stars near SN~2004S must
be quite robust.  On other occasions we determined the photometric
zeropoints using the field star photometry listed in Table
\ref{standards}.  The color terms were determined from observations of
the \citet{Lan92} standards on photometric nights.

The \ubvri light curves of SN~2004S are shown in Fig. \ref{ubvri}, and the
\yjhk light curves of SN~2004S are shown in Fig. \ref{yjhk}. The light
curve templates shown in these figures are derived from the
photometry of SN~2001el \citep{Kri_etal03}. We simply
adjusted the light curve templates in the Y-direction to minimize the
$\chi^2$ statistic of the fits to SN~2004S.\footnote[14]{For \bvri
we used cubic splines to fit the SN~2001el light curves.  For $UJHK$ we used
polynomials.  The $U$-band photometry of SN~2001el from 29 to 65 days after
the time of $B$-band maximum was synthetic photometry based on $HST$ spectra
kindly made available by P. Nugent.}

We note that because the optical and IR photometry of SN~2001el was
obtained with different telescopes and different filters, the $BV$
photometry of \citet{Kri_etal03} required correction to a uniform
photometric system.  We adopted the filter system of
\citet{Bes79, Bes90}. While the ``method of S-corrections'' 
is not perfect \citep{Str_etal02}, it resolves the systematic differences
of ANDICAM $B$ and $V$ photometry compared to SN data obtained with
the CTIO 0.9-m telescope.  Graphical and tabular values of the S-corrections
can be found in \citet{Kri_etal03} and \citet{Kri_etal04c}.  Officially,
each SN should have its own set of S-corrections.  While many Type Ia SNe
have similar spectra, they are not identical, so applying the S-corrections
adds scatter to the data.  However, since the S-corrections we used were
based on SNe 1999ee and 2001el, and SN~2001el is so similar photometrically
to SN~2004S, correcting SN~2004S data in the same manner as we corrected data of 
SN~2001el may add only $\pm$ 0.01 mag of scatter.

Similary, we calculated IR S-corrections to place the ANDICAM $JHK$
photometry on the system of \citet{Per_etal98}.  We found that corrections
to the $R$- and $I$-band photometry actually made the photometry of
SN~2001el from different telescopes more discordant, not less.  Here we use
only the ANDICAM photometry of SN~2001el (without S-corrections) to produce
$R$- and $I$-band templates for a direct comparison of photometry of SNe
2001el and 2004S.\footnote[15]{In 2001 ANDICAM was used with the CTIO 1.0-m
Yale-AURA-Lisbon-Ohio (YALO) telescope, while beginning in 2003 it has been
used on the 1.3-m ex-2MASS telescope.  We assume that the optics of the
primary and secondary mirrors of the two telescopes are sufficiently similar
that no serious photometric differences result.  We also note that all of
the $U$-band photometry of SN~2001el was obtained with the CTIO 0.9-m and 1.5-m
telescopes, {\em not} with ANDICAM.} The key point is that we
have made the same corrections to the photometry of the two SNe 
in the same way, filter by filter, so that a direct comparsion of the 
light curves of the two objects is consistent.  

The one exception is the $U$-band.  On the one hand, the shapes of the
$U$-band light curves of SNe 2001el and 2004S are quite similar.  On the
other hand, the $U$-band photometry of SN~2004S is ``too faint'',
on average, compared to SN~2001el.  We can use a
single function to fit the \bvrijhk data sets (see Fig. \ref{dmag} below),
but the mean difference of the $U$-band data is discrepant by 0.26 $\pm$ 0.07 
mag. \citet[][Fig. 9]{Jha_etal06a} have shown that normal Type Ia SNe at
maximum light have $B-V \approx -0.1$, but that the $U-B$ colors of these 
objects range more than half a magnitude.  They emphasize that
``this is not an artifact of the reddening correction, nor can it be
explained by variation in the extinction law in these external galaxies.''
Their data imply that ``there is a significant intrinsic dispersion in
$U$-band peak brightness even after accounting for variations in light-curve
shape.''  Since the ultraviolet portion of the spectra of Type Ia SNe
contains many metallic lines, a significant fraction of the dispersion
of $U$-band photometry from object to object must be due to the metallicity
of the progenitors \citep{Hoe_etal98, Pod_etal06}.
Suffice it to say that a full understanding of the $U$-band 
photometry and deductions that may be made from it are beyond the scope
of the present paper.

A visual inspection of Figs. \ref{ubvri} and \ref{yjhk} demonstrates the
remarkable similarity of the light curves of SNe 2001el and 2004S.  Any
differences are more obvious if we plot the residuals of the fits, as
shown in Figs. \ref{opt_res} and \ref{ir_res}. Using SN~2001el templates,
the reduced $\chi^2$ values of the fits to the CTIO 1.3-m data of SN~2004S
are 1.24, 1.67, 1.12, 1.50, and 3.45 for $U$, $B$, $V$, $R$, and $I$,
respectively. For $J$, $H$, and $K$ the reduced $\chi^2$ values of the
fits are 1.90, 5.62, and 0.99, respectively. 

%At $t$ = +26 days SN~2004S is
%brighter than the adjusted SN~2001el templates in all optical filters by
%$\sim$ 0.05 mag. This is to say that it had a slightly brighter secondary
%maximum.  At $t \approx$ +40 days SN~2004S was brighter by more than 0.1
%mag in $U$ compared to our late-time synthetic photometry of SN~2001el
%based on $HST$ spectra.

Very little $Y$-band photometry of SNe has been published so far.  
Previously, we published some $Y$-band data of SNe 1999ee and 2000bh
\citep{Kri_etal04b} and noted that the second maximum of SN~2000bh was
brighter than the first maximum. SN~2004S has shown the same behavior.  
\citet{Ham_etal06} indicate that many Type Ia SNe observed
at Las Campanas Observatory as part of the Carnegie Supernova Project also
show stronger second maxima in the $Y$-band compared to the first 
maxima.\footnote[16]{Some preliminary light curves can be seen at
http://csp1.lco.cl/~cspuser1/CSP.html.}

The origin of the secondary maximum in the light curves of Type Ia SNe
has recently been discussed by \citet{Kas06}.  His results derive from
1-D (spherically symmetric) modelling.  The basic idea is that 
a mixture of 75 percent $^{56}$Co, 24 percent iron, and 1 percent $^{56}$Ni,
has a near-IR emissivity that sharply peaks at a temperature of 7000 K.  This 
corresponds to the composition and temperature of the ejecta of
a Type Ia SN 40 days after explosion, or 20 days after T($B_{max}$).
(The exact composition is not critical, however, so long as we are considering
iron group elements.)
At this time the iron group elements undergo a transition from a state
of double to single ionization.  This in turn causes a redistribution 
of energy from optically thick shorter wavelengths to
optically thin near-IR wavelengths.  If the $Y$-band secondary
maximum is brighter than the first maximum in most or all cases, the process which
gives rise to the secondary maximum must be happening to the greatest degree
at 1.03 $\mu$m.  

\citet{Kas06} also suggests that a transition from singly ionized iron group elements
to neutral atoms, occurring some 80 days after T($B_{max}$), may give rise to a {\em
third} maximum.  In Figs. \ref{yjhk} and \ref{vh_all} we see evidence for this
phenomenon in the light curves of SN~2004S, at the 2.8$\sigma$ level in the $J$-band
and up to 5.3$\sigma$ in the $H$-band. There exist very few near-IR measurements of
Type Ia SNe at these late epochs. We now have an extra movitation to observe these
objects in the near-IR at late epochs $-$ to check Kasen's prediction.

Using the light curve fitting code of \citet{Pri_etal06}, which relies on \bvri data, we find
for SN~2004S that T($B_{max}$) = JD 2,453,039.87 $\pm$ 0.25, \dmm = 1.14 $\pm$ 0.01, and
E($B-V$)$_{host}$ = 0.00 $\pm$ 0.01.  The derived value of the decline rate is statistically
equal to that of SN~2001el, \dmm = 1.13 $\pm$ 0.04 
\citep{Kri_etal03}.\footnote[17]{In the context of the MLCS2k2 system \citet{Jha_etal06b}
give $\Delta$ = $-$0.19 $\pm$ 0.03 mag for SN~2001el.  Using a copy of that software we
obtain $\Delta$ = $-$0.24 $\pm$ 0.04 mag for SN~2004S.  $\Delta < 0$ is a measure
of the overluminosity of these SNe compared to the fiducial object of the system. }  
The \dmm\ solution confirms that SN~2004S had minimal host galaxy
reddening, A$_V$(host) $\approx$ 0.00 $\pm$ 0.03. A more robust solution for the host galaxy
extinction is obtained using the optical and IR data (see below). The \dmm\ method gives a
distance modulus of $m-M$ = 33.32 $\pm$ 0.17 mag on an $h$ = 0.72 scale
\citep{Fre_etal01}.\footnote[18]{We define $h \equiv$ H$_0$ / (100 km s$^{-1}$ 
Mpc$^{-1}$).}
This is to be compared with $m-M$ = 33.07 $\pm$ 0.21 mag from the recession velocity in the
CMB frame.

Using the SN~2001el $V$-band template, we estimate that the $V$-band maximum of
SN~2004S is $V_{max}$ = 14.48 $\pm$ 0.03.  With A$_V$(tot) = 0.38 mag (see \S4)
and adopting a distance modulus of $m-M$ = 33.07 $\pm$ 0.21 mag from the radial
velocity in the CMB frame and Hubble's Law, we obtain M$_V$(max) = $-$18.97 $\pm$
0.22 mag. The expected value for a Type Ia SN of average decline rate is M$_V$ =
$-$19.13 on an $h$ = 0.72 scale \citep{Phi_etal99}.  Thus, SN~2004S has an
absolute $V$ magnitude at maximum within 1-$\sigma$ of the nominal value for its
decline rate.

Using the SN~2001el templates, we estimate that the S-corrected, K-corrected
apparent magnitudes at maximum light of SN~2004S were $J_{max}$ = 14.806,
$H_{max}$ = 14.975, and $K_{max}$ = 14.766, with uncertainties of $\pm$ 0.03 mag.  
The total extinction due to dust in our Galaxy and in the host of SN~2004S is
A$_J$ = 0.107, A$_H$ = 0.068, and A$_K$ = 0.043 mag, with uncertainties of $\pm$
0.02 mag or less (see \S4). Adopting a distance modulus of $m-M$ = 33.07 $\pm$ 0.21 mag
from the radial velocity in the CMB frame and Hubble's Law, we obtain IR absolute
magnitudes at maximum light of M$_J$ = $-$18.37, M$_H$ = $-$18.16, and M$_K$ =
$-$18.35, with uncertainties of $\pm$ 0.22 mag.  These are to be compared with
the mean values of more than 20 objects with \dmm\ $\lesssim$ 1.8 of M$_J$ =
$-$18.61, M$_H$ = $-$18.28, and M$_K$ = $-$18.44 \citep[][Table 17]{Kri_etal04c}.  
Taking the numbers at face value, SN~2004S is slightly underluminous in the IR,
which contradicts the implication from MLCS2k2 analysis that the object is 
somewhat overluminous.  In any case, the near-IR absolute magnitudes are
still consistent with the notion that Type Ia SNe with \dmm\
$\lesssim$ 1.8 are standard candles in the near-IR \citep{Kri_etal04a}, with an
intrinsic dispersion of roughly $\pm$0.15 mag.

For SN~2004S \citet{Mis_etal05} found A$_V$(tot) = 0.542 $\pm$ 0.167 mag.  This
implies host galaxy extinction of A$_V$(host) $\approx$ 0.23 mag, with a rather
large uncertainty.  These authors adopt a distance modulus of 32.94 mag on an
$h$ = 0.65 scale, equivalent to $m-M$ = 32.72 mag on an $h$ = 0.72 scale.  They
used a somewhat lower heliocentric redshift for the host galaxy (2730 km
s$^{-1}$) and corrected for the Virgocentric infall of the Local Group, but do
not quote a velocity in the CMB frame.  One can also use the flow model of
\citet{Ton_etal00} to give an estimate of the distance to SN~2004S.  Along a
vector towards the direction of a galaxy in the sky, one determines the distance
at which the flow model gives a radial velocity that matches the observed
heliocentric velocity.  For the host of SN~2004S that distance is 36.25 Mpc on an
$h$ = 0.784 scale, using the \citet{The_etal05} heliocentric radial velocity of
2808 km s$^{-1}$.  On an $h$ = 0.72 scale the equivalent distance modulus is
$m-M$ = 32.98 mag, statistically in agreement with the value of 33.07 $\pm$ 0.21
mag from the CMB velocity and Hubble's Law.

Since SN~2001el is known as one of the few Type Ia SNe to exhibit
measurable polarization, implying an asymmetrical explosion
\citep{Wan_etal03b}, SN~2004S may also have exhibited the same effect.
See \citet{Cho_etal06} for a discussion.

A final topic to mention in this section is the late-time photometry given
in Tables \ref{opt_photom} and \ref{ir_photom}.  As \citet{Sol_etal04}
found in the case of SN~2000cx, six months to a year after T($B_{max}$)
SN~2004S continued to get fainter at optical wavelengths, 
but in the near-IR it reached a plateau and stayed constant.  
This confirms that at late times the near-IR flux of a Type Ia SN 
becomes a larger and  larger fraction of the total flux 
\citep[see Fig. 13 of][]{Sol_etal04}.  We show in Fig. \ref{vh_all} 
our $V$- and $H$-band photometry of SN~2004S.  The $B$-band photometry
from 50 to 410 days after T($B_{max}$) gives dm/dt = 1.50 $\pm$ 0.01
mag per 100 days.  The late-time photometry ($t >$ 196 d) in the 
$V$, $R$ and $I$ bands gives dm/dt = 1.42 $\pm$ 0.04,
1.55 $\pm$ 0.12, and
1.40 $\pm$ 0.10 mag per 100 days, respectively.  \citet{Sol_etal04} 
found decline rates of $\sim$1.4 mag per 100 d for the optical 
light curves of SN~2000cx.  Stritzinger \& Sollerman (2007, in preparation)
present late-time photometry of SN~2001el.

At late times the optical minus IR color of SN~2004S became very red.
$V-H = +0.27 \pm 0.10$ at $t$ = 76 d (see Figs. \ref{vir_temp8}
and \ref{vir} below).  At $t$ = 342 d the $V-H$ color was one magnitude redder
(+1.27 $\pm$ 0.20), and at $t$ = 416 d it was one magnitude redder still
(+2.27 $\pm$ 0.20).

\section {Optical and Infrared Spectra of SN~2004S}

In Table \ref{spectra} we summarize some basic information about the
spectra of SN~2004S presented here. The spectra were obtained with the
CTIO 1.5-m telescope, the 3.58-m Telescopio Nazionale Galileo (TNG), the
2.56-m Nordic Optical Telescope (NOT), and the 10-m Keck telescope, and were
reduced following the standard recipe for longslit spectroscopy.
Spectropolarimetry of SN~2004S, obtained with the Keck telescope, and a
comparison to SN~2001el, are also given by \citet{Cho_etal06}.

In Fig. \ref{04S_stack} we show optical spectra of SN~2004S obtained
between 1.7 and 42.5 days after T($B_{max}$).
In Fig. \ref{spectral_comp} we compare the optical spectra of SN~2004S
at three epochs to several other normal Type Ia supernovae: SN~1996X \citep{Sal_etal01},
SN~1998bu \citep{Her_etal00}, SN~1999ee \citep{Ham_etal02}, SN~2001el 
\citep{Wan_etal03b}, and SN~2003du \citep{Gera_etal04, Sta_etal06}.
Clearly, the spectral evolution of SN~2004S closely follows that of
other normal Type Ia SNe. It is worth noting that the largest differences between
the SNe are observed in the Ca~II H\&K and infra-red triplet lines.
Otherwise, all the spectra are nearly identical.

In Fig. \ref{comp_at_max} one can see how remarkably similar SNe 2001el
and 2004S are at maximum light.
\citet{Bra_etal06} investigated the maximum-light spectra of 24 Type Ia
SNe. Their ``core-normal'' group of 7 objects included SNe 1998bu and
2001el.  They noted the high-velocity Ca II absorption of SN~2001el and
also identified the three local minima in the spectrum between 4600 and
4900 \AA\ as being due to high-velocity Fe II.
Since SN~2004S shows the
same features, the identification undoubtedly holds for SN~2004S as   
well.  \citet{Bra_etal06}  also noted
that SN~2001el had the strongest high-velocity Fe II of all
``core-normal'' SNe. In Fig. \ref{comp_at_max} one can see that
high-velocity Fe\,II features in SN~2004S are as strong as in SN~2001el, thus
strengthening the similarity between the two SNe. Besides, the two SNe had
equal Si~II ratios, $\mathcal{R}$(Si\,II) $\approx 0.3$, as defined by 
\citet{Nug_etal95}.

Despite the similarities between SNe 2001el and 2004S, there {\em are} 
a few spectroscopic differences. The depth of the Si II absorption
observed at 6150 \AA\ is weaker in SN~2004S. Part of the blend due to Fe
II just blueward of 5000 \AA\ is different in the two SNe.  The greater
dust extinction along the line of sight to SN~2001el produces some
absorption due to the Na D lines at 5889/5896 \AA.  SN~2001el had a
slight shoulder in its Ca II absorption at 7900 \AA\, whereas SN~2004S
did not.

 Larger differences are observed in the Ca\,II infra-red triplet
(Ca\,II~IR3). The differences may be attributed to the presence of high-velocity 
features (HVFs) approximately centered at 20000~km\,s$^{-1}$.
In Fig. \ref{vel_comp} we see a comparison of the velocity profiles
of the Ca\,II~IR3 in SNe 1999ee, 2001el, and 2004S. The line profiles are  
normalized to a local pseudo-continuum, approximated by a straight line
in a way similar to the definition of the Si~II ratio \citep[see][]{Nug_etal95}.
Around T($B_{max}$) the Ca\,II HVFs in SNe 2001el and 2004S have equal strength,
but at the later epochs they are stronger in SN 2004S. At the same time the photospheric
Ca\,II line is stronger in SN 2001el. The HVFs in SN 1999ee at the time of
maximum light are weaker
than in SNe 2001el and 2004S, but are stronger 20 days after.
Note, however, that while SN~1999ee also had high-velocity Ca II, it was placed
by \citet{Bra_etal06} in their ``shallow-silicon'' category; it
had weaker Si II and S II features than SN~2001el.  We note that
SN~1999ee had a much slower decline rate \citep[\dmm = 0.94 $\pm$
0.06;][]{Str_etal02}, which, along with the shallower absorption lines
of singly ionized species, is consistent with it having been a hotter
explosion than SNe 2001el and 2004S.

Another difference emerges when comparing the velocity inferred from the
minimum of the Si\,II~$\lambda$6355 line. Around T$(B_{max}$) SN 2004S had a velocity of
$\sim9300$~km\,s$^{-1}$, while in SN 2001el it was $\sim10200$~km\,s$^{-1}$.
Analyzing the spectrophotometric properties of a sample of Type Ia SNe,
\citet{Ben_etal05} found that they can be
statistically divided into three groups mainly on the behavior of the rate
of decrease of the expansion velocity of the Si\,II~$\lambda$6355 absorption
after the time of maximum light, 
$\langle\dot{v}\rangle$ (`velocity gradient'), and also according to the
velocity of the Si\,II~$\lambda$6355 at T($B_{max}$), $v_{max}$(Si\,II). Respectively,
the groups have been called high velocity gradient (HVG;
$\langle\dot{v}\rangle = 97\pm16$ km\,s$^{-1}$d$^{-1}$, $v_{max}$(Si\,II) = 12200$\pm1100$),
low velocity gradient (LVG;
$\langle\dot{v}\rangle = 37\pm18$ km\,s$^{-1}$d$^{-1}$,  $v_{max}$(Si\,II) = 10300$\pm300$),
and FAINT ($\langle\dot{v}\rangle = 87\pm20$ km\,s$^{-1}$d$^{-1}$,   $v_{max}$(Si\,II) = 
9200$\pm600$).
The FAINT group includes SNe that are intrinsically dim, on average $\sim 2$ mag
fainter than SNe belonging to the other two groups. SN 2001el had $\dot{v}=31\pm5$
km\,s$^{-1}$d$^{-1}$,
and \citet{Ben_etal05} placed it in the LVG group. For SN 2004S, 
however, we measure  $\dot{v}=69\pm5$ km\,s$^{-1}$d$^{-1}$, which is in between the LVG 
group on one hand, and HVG and FAINT groups on the other. Along with the relatively
low velocity at the time of maximum light, this suggests that SN~2004S may be an 
intermediate object in the context of the analysis of the Si\,II~$\lambda$6355 
line velocity evolution.

A single IR spectrum of SN~2004S was obtained 15 days after T($B_{max}$)
using the Near-Infrared Camera and Spectrograph at the TNG. An Amici prism
was used as a disperser, providing the whole near-IR spectral range in one
exposure but with very low resolving power of $\sim100$.
In Fig. \ref{ir_comp} we compare the near-IR spectra of SNe 2004S, 1999ee,
and 1998bu; the near-IR spectra of these three objects are dominated
by the same singly ionized species, mostly of iron group elements.
\citet{Mar_etal03} showed that the abrupt change of the flux near
1.52 $\mu$m (rest wavelength $\sim1.57 \mu$m) observed in many Type Ia SNe
defines the transition from partial to complete silicon burning.
In SN 2004S we measure a velocity of $\sim11000$\,km\,s$^{-1}$, which is
consistent with the measurements of other SNe and the model predictions for
a normal-bright Type Ia SNe \citep{Mar_etal03}.

\section {Considerations of reddening and extinction of SNe 2001el and 2004S}

The $V-J$, $V-H$, and $V-K$ color curves of SN~2004S are shown in Fig.
\ref{vir_temp8}. Using our zero-reddening loci for Type Ia SNe with mid-range
decline rates, we find total color excesses (i.e. due to Galactic dust and
host galaxy dust) of E($V-J$) = 0.324 $\pm$ 0.060, E($V-H$) = 0.371 $\pm$
0.056, and E($V-K$) = 0.316 $\pm$ 0.030 mag.  We note that the
$V-K$ curve is by far the best match as to shape.  To scale the color
excesses to estimates of A$_V$, the appropriate coefficients are 
1.394 $\pm$ 0.110, 1.217 $\pm$ 0.058, and 1.130 $\pm$ 0.029, respectively.  These
coefficients differ slightly from the values in Eqs. 2, 3, and 4 of
\citet{Kri_etal04b}.  The \citet{Car_etal89} values of A$_{\lambda}$/A$_V$
are appropriate for determining extinction of stars with normal spectra.
Since SNe have such different spectra we used the values in Table 8
of \citet{Kri_etal06}, which are based on spectra of SN~1999ee at maximum 
light.  The uncertainites assigned to these coefficients correspond to a 20 percent
uncertainty in A$_{\lambda}$/A$_V$.  For SN~2004S we obtain three estimates of A$_V$,
namely 0.452 $\pm$ 0.090, 0.452 $\pm$ 0.071, and 0.357 $\pm$ 0.035
mag. The weighted mean is A$_V$(tot) = 0.384 $\pm$ 0.030 mag, of which 0.313 $\pm$ 
0.031 mag is due to dust in our Galaxy.  It follows that A$_V$(host) = 
0.071 $\pm$ 0.043 mag, implying a finite but small amount of host galaxy
extinction.  This is consistent with the host galaxy extinction derived using
the \dmm\ method.

Since we have already demonstrated that the optical and IR light curves of
SNe 2001el and 2004S are so similar, it is no suprise that the shapes of the
photometric color curves of the two objects are also nearly identical.  
By adjusting a color template from SN~2001el to the corresponding color
curve of SN~2004S we can determine the difference of the colors
and therefore the difference of the color excesses.  In Fig. \ref{vir}
we show the same SN~2004S data as in Fig. \ref{vir_temp8} but with the color
templates derived from data of SN~2001el, adjusted to minimize the $\chi^2$
statistic.  We find that we need to shift the SN~2001el templates by
$\Delta$($V-J$) = $-$0.144 $\pm$ 0.044, 
$\Delta$($V-H$) = $-$0.175 $\pm$ 0.035, and 
$\Delta$($V-K$) = $-$0.187 $\pm$ 0.068 mag. 

We may next correct for the effect of Galactic dust.  SN~2001el has
E($B-V$)$_{Gal}$ = 0.014, while SN~2004S has E($B-V$)$_{Gal}$ = 0.101
\citep{Sch_etal98}.  Thus, the difference of the Galactic components of $V$-band
extinction toward these two objects is (0.101 $-$ 0.014) times R$_V$ of 3.1, or
0.270 $\pm$ 0.027 mag.  The differences of the color excesses due to dust in our 
Galaxy are:
$\Delta$E($V-J$)$_{Gal}$ = 0.270 / 1.394 = 0.194 $\pm$ 0.019,
$\Delta$E($V-H$)$_{Gal}$ = 0.270 / 1.217 = 0.222 $\pm$ 0.022, and
$\Delta$E($V-K$)$_{Gal}$ = 0.270 / 1.130 = 0.239 $\pm$ 0.024 mag.
We find that the
host galaxy dust of NGC 1448 reddened SN~2001el more than the host
galaxy dust of MCG -05-16-021 reddened SN~2004S by these amounts:
$\Delta$E($V-J$)$_{host}$ = 0.194 + 0.144 = 0.338 $\pm$ 0.048, 
$\Delta$E($V-H$)$_{host}$ = 0.222 + 0.175 = 0.397 $\pm$ 0.041, and 
$\Delta$E($V-K$)$_{host}$ = 0.239 + 0.187 = 0.426 $\pm$ 0.072 mag.  
To obtain the difference
of host galaxy $V$-band extinction, we need to scale these values
by coefficients appropriate to the dust in the two host galaxies.
\citet{Kri_etal03} found A$_V$(tot) = 0.57 $\pm$ 0.05 for SN~2001el.
With A$_V$(Gal) $\approx$ 0.043 towards SN~2001el, A$_V$(host) $\approx$
0.53 mag.  Thus, SN~2001el suffered approximately
an order of magnitude (0.53/0.07 $\approx$ 7.6) more host galaxy extinction 
compared to SN~2004S. For SN~2001el \citet{Wan_etal03b} give R$_V$ = 2.88 $\pm$ 
0.15 from the peak wavelength of the interstellar polarization.\footnote[19]{In
our view the polarization curve shown in Fig. 8 of \citet{Wan_etal03b} 
does not have a well constrained maximum, so their value of R$_V$ should probably  
be assigned a larger numerical uncertainty.}  \citet{Jha_etal06b}
give R$_V$ = 2.40 $\pm$ 0.19 for the host galaxy dust of SN~2001el.
Using the values in Table 8 of \citet{Kri_etal06} and this range of
values of R$_V$, we find that the difference of host galaxy $V$-band
extinction of the two SNe is A$_V$ = 0.471 $\pm$ 0.027 mag.  Adding
this to the magnitude difference of the $V$-band light curves, corrected
for Galactic extinction (column 5 of Table \ref{adjust} below), the
implied difference of the distance moduli of the two objects is
1.936 $\pm$ 0.044 mag.

It is not absolutely necessary to make any a priori assumptions about the
parameters of interest here.  In Table \ref{adjust} we lay out a spread sheet
relating to the implied dust extinction in the hosts of SNe 2001el and 2004S.  
In column 2 we give the magnitude offsets necessary to minimize the $\chi^2$
statistic of fitting the SN~2001el light curve templates to SN~2004S.  
In column 5 of Table \ref{adjust} we have corrected the values of
column 2 for the extinction due to Galactic dust.

In Fig. \ref{dmag} we show the values from column 5 of Table \ref{adjust} vs. the
wavelengths of the photometric filters.  As the wavelength tends to infinity, the
effect of the dust in the two host galaxies becomes negligible.  In other words,
in the absence of host galaxy dust extinction, all the points (excepting $U$)
would lie along some horizontal line near the top of the graph.  This is most
directly interpreted as the difference of the distance moduli ($\equiv \Delta\mu$)  
of the two objects.\footnote[20]{Because Type Ia SNe exhibit an intrinsic
dispersion at any given decline rate of $\pm$ 0.15 mag, two objects with identical
explosion mechanisms, or at least producing the same amount of $^{56}$Ni, would
not necessarily give the exact same extinction-corrected absolute magnitudes.  
And even if the two objects produced the same number of ergs per second, the
explosions could be asymmetric.  One elongated fireball might be observed end-on,
while the other could be observed side-on, giving different apparent
brightnesses.}

We have carried out a multi-dimensional $\chi^2$ minimization, to find the values
of R$_V$, A$_V$ and $\Delta\mu$ that give the best match to the values in column
5 of Table \ref{adjust}.  See Table \ref{extinction} and Fig. \ref{contour}. We
find R$_V$ = 2.15$^{+0.24}_{-0.22}$, A$_V$ = 0.472 $\pm$ 0.025 mag, and
$\Delta\mu$ = 1.936 mag.  Note the close agreement of A$_V$ and $\Delta\mu$ with the
values obtained from a consideration of the $V-$[$JHK$] color curves of
SNe 2001el and 2004S.

Using only \bvri data we find R$_V$ = 2.51$^{+0.50}_{-0.36}$.  This is
statistically in agreement with the values of \citet{Jha_etal06b} and
\citet{Wan_etal03b} for the host galaxy dust of SN~2001el.  However, our value
derived from \bvrijhk data is considerably smaller than that of
\citet{Wan_etal03b}, derived from polarization data.  Our ``best'' value
implies that the dust in the host of SN~2001el had an R$_V$ value considerably
less than the canonical Galactic value of 3.1.

In Fig. \ref{contour} we show the 1-$\sigma$ contours of our extinction solutions
using 4, 5, and 7 filter photometry.  As the number of filters and the wavelength
range of the filters increase, the contours become smaller and rounder.  In 
Table \ref{extinction} one can see how the uncertainties decrease as the
number of filters increases from 4 to 6.  Adding the $K$-band data changed the
solution only slightly.  This is undoubtedly due to the larger uncertainty
of the magnitude shift necessary to fit the SN~2004S $K$-band data with the 
corresponding SN~2001el template.

We can estimate the total $V$-band extinction of SN~2001el as follows. The
extinction due to Galactic dust along the line of sight is A$_V$(Gal) = 0.043 $\pm$
0.004 mag.  The host galaxy $V$-band extinction of SN~2001el is 0.472 $\pm$ 0.025
mag (difference of SNe 2001el and 2004S) plus the host galaxy extinction of SN~2004S
(0.071 $\pm$ 0.043 mag from $VJHK$ data, but possibly zero if we rely on the \dmm\
solution using \bvri data). For SN~2001el A$_V$(tot) = 0.586 $\pm$ 0.050 mag.  
This compares very well with our previously published value of 0.57 $\pm$ 0.05 mag
\citep{Kri_etal03}.

\section{Modelling distance uncertainty and dust}

For Type Ia SNe to be useful as accurate probes for cosmology
it is important that systematic errors be minimized.
Future probes of Dark Energy properties will need to control
systematic uncertainties to better than a few percent to
provide useful constraints (Miknaitis et al. 2006, in preparation).
For a typical E($B-V$) $\approx$ 0.10 mag, a systematic error of 0.2 in
R$_V$ leads to a systematic error of 0.02 mag in the distance modulus.
The range of extinction properties seen in nearby supernovae,
such as SN~2004S, suggests that it is important to measure
the extinction law to avoid making a systematic error in
distance determination. But our photometry of SN~2004S shows how
difficult it is to precisely constrain $R_V$ using just
four optical filter bands.

To better understand how well dust extinction can be corrected
in a sample of well-observed Type Ia SNe, we have simulated the
process used to recover $R_V$, $A_V$ and $\Delta\mu$ for
SN~2004S. In the simulation we have assumed the following:
\begin{itemize}

\item Supernovae have a probability distribution for $A_V$ like
 that found by \citet{Jha_etal06b}. This is an exponential 
 probability distribution function (pdf) peaking at 
 zero extinction and having a scale factor of 0.4 mag.

\item $R_V$ has a Gaussian pdf with a mean value of 2.4 and
standard deviation of 0.4.

\item The extinction as a function of bandpass is given by
 A$_i$/A$_V \; =\; a_i + b_i$/R$_V$ where $i$ corresponds to one
of the \bvrijhk bands and the coefficients $a_i$ and $b_i$ are
given in Table 8 of \citet{Kri_etal06}.

\end{itemize}

We then did a Monte Carlo simulation selecting pseduo-random
values of $R_V$ and $A_V$ which define the extinctions in the
\bvrijhk bands. How accurately these input parameters can be 
recovered depends on two factors: the number of filters used
in the observations and the signal-to-noise ratio of the
observations. The simulations were done assuming observations
in four bands ($BV\hspace{-0.5mm}RI$), 
five bands ($BV\hspace{-0.5mm}RIJ$), and all seven
filters ($BV\hspace{-0.5mm}RIJHK$). The signal-to-noise ratio in each
band was simulated by setting a variance, $\sigma^2_{phot}$,   
about the true extinction. Each realization
of the Monte Carlo simulation selected a random value     
that had a Gaussian distribution with that variance and   
centered on the true bandpass extinction.

The resulting simulated measurements were then fit with three
free parameters, $R_V$, $A_V$, and $\Delta\mu$, using a $\chi^2$  
minimization technique. $\Delta\mu$ is the offset in the recovered
distance modulus and the measure for how well the various filter set
and photometric uncertainties perform.

A sample result for 30,000 events with a photometric error
of 0.02 mag and recovery in \bvrij bands is shown in Fig. \ref{scat}. The
difference between the input A$_V$ and recovered A$_V$ (i.e. the A$_V$ 
error) is plotted along with the corresponding
R$_V$ error to show the correlation between these parameters.  Note that
on average the recovered values of R$_V$ and $A_V$ have no significant 
systematic error.

In Fig. \ref{mu} we explore how adding IR bands improves the distance 
estimation.  As expected, the use of \bvrijhk comes closest to the 
optimal recovery of the distance modulus. The
use of only \bvri bands means that the recovered distance uncertainty
rises sharply with photometric error. For \bvri, photometric
errors of only 0.03 mag result in distance uncertainites of 0.2~mag,
which is the same order as the intrinsic scatter in the luminosities
of Type Ia SNe.  We note that our simulation was quite simple and assumes  
that the true color distribution for unextincted Type Ia SNe at
all decline rates is known with perfect precision.

The addition of the rest frame $J$-band to the filter complement
decreases the distance uncertainty by a factor of 2.7.
The use of rest frame $J$-band observations greatly improves
the precision in estimating the distance modulus when independently
fitting a two-parameter ($R_V$, $A_V$)
extinction law.  Adding $H$ and $K$ further 
improves the uncertainty of the distance modulus, but only by an 
additional 30 percent.

Any future SN survey to be carried out with a satellite-borne telescope
would wisely observe Type Ia SNe in rest frame optical and IR bands.  
To determine the cosmic equation of state parameter ($w$) to within 10
percent requires controlling photometric systematic errors at the 0.01
to 0.02 mag level (Miknaitis et al., 2006, in preparation).  At minimum
we must know the appropriate mean value of R$_V$ for high redshift SNe.  
Better still would be the option to determine R$_V$ accurately for a 
large fraction of those yet-to-be discovered objects.

\section{Conclusions} 

In this paper we provided optical and IR photometry of the normal Type Ia
SN~2004S.  Using optical and near-IR data, it was found to have minimal 
host galaxy extinction, with A$_V$(host) = 0.07 $\pm$ 0.04 mag.  

There are remarkable similarities of the shapes of the optical and IR light
curves of SNe 2004S and 2001el.  This is the first time we have seen two objects
exhibit such similar $JHK$ light curves at the time of maximum light until 65 
days afterwards.  We might assume that the two objects had identical explosion
mechanisms, or at least produced the same amount of $^{56}$Ni. Given that Type Ia
SNe exhibit an intrinsic dispersion of roughly $\pm$0.15 mag at any given decline
rate, SNe 2001el and 2004S could have produced identically shaped
filter-by-filter light curves, yet could have had absolute magnitudes that differ
by 0.15 mag or more.  For our purposes it is reasonable to
interpret the difference of their extinction-corrected magnitudes in any band
(except $U$) as a measure of the difference of their distance moduli.  More 
generally, we may assume that they had identical extinction-corrected colors.

SNe 2001el and 2004S not only had similar light curves, but very similar
optical and near-IR spectra, exhibiting high-velocity absorption due to singly
ionized Ca and Fe.  Except for the high velocity features in their ejecta prior to $t
\approx$ +14 d, they can be considered spectroscopically normal \citep{Bra_etal06}.

The similarities of the light curves of SNe 2001el and 2004S allow us to use the
photometry, filter by filter, to derive the {\em difference} of the host galaxy
extinctions without making any a priori assumptions about the unreddened
colors of Type Ia SNe, values of R$_V$, or the reddening law between the $K$-band
and infinite wavelength.  Since SN~2001el had roughly an order of magnitude more
host galaxy extinction compared to SN~2004S, our multi-dimensional $\chi^2$
minimization primarily tells us something about the dust in the host of SN~2001el.  
For this object we find R$_V$ = 2.15$^{+0.24}_{-0.22}$ and 
A$_V$(tot) $\approx$ 0.59 $\pm$ 0.05 mag.  Because of the
advantages of combining optical and IR photometry to determine extinction, it is
important that we know the shapes and zeropoints of the color curves of Type Ia
SNe.  Our analysis gives us unreddened solutions for SNe 2001el and 2004S.

The $V-K$ color curves of SNe 2001el and 2004S have the same shape as
our template based on the not-so-well sampled light curves of 8 mid-range
decliners.  Our $V-J$ and $V-H$ templates do not have quite the same
shapes as the color curves of SNe 2001el and 2004S.  In the near future
it should be possible to construct more robust color curves from a larger
database of objects, which will reflect more realistically the inherent
scatter exhibited by Type Ia SNe.

We find that SN~2004S had $VJHK$ absolute magnitudes at maximum light within
1-$\sigma$ of the mean of other Type Ia SNe of comparable decline
rate. Any systematic differences may only be due to the unknown peculiar
velocity of the host galaxy.  The $JHK$ absolute magnitudes at maximum of
SN~2004S are consistent with the notion that in the near-IR Type Ia SNe
(excepting the fastest decliners) are standard candles.

We can derive host galaxy extinction much more accurately using optical and IR
photometry than using optical photometry alone, as we have shown graphically and
quantitatively.  Our Monte Carlo simulations indicate that rest frame \bvrij
photometry gives distance uncertainties 2.7 times smaller than just using rest
frame \bvri photometry.  Assuming a fixed value of R$_V$ for high-redshift
SNe runs the risk of systematic errors if the fixed value is far from the 
true mean.  Here we showed that adding one additional rest frame band in the
near-IR substantially improves the resulting uncertainty of determining
the distance moduli of Type Ia SNe.  Rest frame $J$-band
corresponds to 3.6 $\mu$m in the observer frame at $z \sim$ 2, which is
observable with space-based telescopes.  Having rest frame optical and near-IR
data for Type Ia SNe will lead to stricter constraints on models of Dark Energy.

\vspace {1 cm}

\acknowledgments

This work is partly based on observations collected at the Nordic
Optical Telescope (NOT) and Italian Telescopio Nazionale Galileo (TNG)
located at the Spanish Observatorio del Roque de los Muchachos of
the Instituto de Astrofisica de Canarias, on the island La Palma.
We would like thank G\"{o}ran \"{O}stlin, Matthew Hayes, Pasi Hakala, Gavin
Ramsay, Else van den Besselaar, Thomas Augusteijn, Sofia Feltzing, Per
Knutsson,  Ingemar Lundstr\"{o}m, Susanne Aalto, Eva Manthey and Eva
Karlsson who gave up part of their NOT time and observed SN~2004S.

The CTIO 1.3-m telescope is operated by the Small
and Moderate Aperture Research Telescope System (SMARTS) Consortium.
We are particularly grateful for the scheduling flexibility of SMARTS.
We made use of the NASA/IPAC Extragalactic Database (NED), the SIMBAD 
database, operated at CDS, Strasbourg, France, and the Two Micron
All-Sky Survey.  We thank Mark Phillips for many useful comments on the paper.

\newpage

\begin{deluxetable}{cccccc}
%\tabletypesize{\scriptsize}
\tablewidth{0pc}
\tablecaption{Optical Photometric Sequence near SN 2004S\tablenotemark{a}\label{standards}}
\tablehead{   \colhead{Star ID\tablenotemark{b}} &
\colhead {$U$} & \colhead{$B$} &
\colhead{$V$} & \colhead{$R$} & \colhead{$I$} } 
\startdata
1 &  16.841 (0.011) & 16.442 (0.008) & 15.612 (0.006) & 15.140 (0.006) & 14.727 (0.005)  \\    
2 &  19.282 (0.066) & 18.504 (0.012) & 17.441 (0.010) & 16.777 (0.015) & 16.222 (0.008)  \\  
3 &  18.134 (0.026) & 18.157 (0.014) & 17.609 (0.010) & 17.257 (0.011) & 16.884 (0.015)  \\
4 &  19.081 (0.054) & 17.897 (0.011) & 16.527 (0.007) & 15.642 (0.005) & 14.845 (0.006)  \\ 
5 &  19.255 (0.051) & 18.556 (0.017) & 17.509 (0.009) & 16.875 (0.006) & 16.325 (0.007)  \\
6 &  16.377 (0.008) & 16.282 (0.008) & 15.590 (0.005) & 15.175 (0.006) & 14.771 (0.006)  \\
7 &  18.206 (0.042) & 17.848 (0.010) & 16.996 (0.007) & 16.501 (0.008) & 16.056 (0.006)  \\
\enddata
\tablenotetext{a} {The numbers in parentheses are 1-$\sigma$ uncertainties (mean errors of
the mean).}
\tablenotetext{b} {The identifications are the same as in Fig. \ref{finder}.}
\end{deluxetable}

\begin{deluxetable}{ccccc}
\tablewidth{0pc}
\tablecaption{Infrared Photometric Sequence near SN~2004S\tablenotemark{a}\label{ir_stds}}
\tablehead{   \colhead{Star ID\tablenotemark{b}} &
\colhead{$Y$} & \colhead{$J$} & \colhead{$H$} & \colhead{$K$} } 
\startdata
1  & 14.450 (0.005) & 14.166 (0.009) & 13.794 (0.009) & 13.682 (0.017) \\
2  & 15.828 (0.010) & 15.448 (0.010) & 14.879 (0.009) & 14.751 (0.039) \\
3  & 16.595 (0.019) & 16.441 (0.023) & 16.117 (0.024) & \nodata  \\
4  & 14.340 (0.005) & 13.918 (0.008) & 13.273 (0.009) & 13.100 (0.017) \\
%
%  the JHK values above include the uncertainties of our only Persson star
%  sigma JHK = 0.007, 0.007, 0.015.  Below we see the uncertainties
%  based on the unweighted avg of the individ measures of star 4 wrt the std
%  and the means of differential of 1,2, and 3 wrt number 4. 
%
%  Since I originally made the table, I found a different Y-band calibration.
%  This requires adding 0.017 mag to the Y-band values. 
%
%1  & 14.433 (0.005) & 14.166 (0.005) & 13.794 (0.006) & 13.682 (0.015) \\
%2  & 15.811 (0.010) & 15.448 (0.007) & 14.879 (0.005) & 14.751 (0.036) \\
%3  & 16.578 (0.019) & 16.441 (0.022) & 16.117 (0.023) & \nodata  \\
%4  & 14.323 (0.005) & 13.918 (0.003) & 13.273 (0.005) & 13.100 (0.007) \\
\enddata
\tablenotetext{a} {The numbers in parentheses are 1-$\sigma$ uncertainties (mean errors
of the mean).}
\tablenotetext{b} {The identifications are the same as those in Table \ref{standards} 
and Fig. \ref{finder}.}
\end{deluxetable}

\begin{deluxetable}{ccccccc}
\tabletypesize{\scriptsize}
\tablewidth{0pc}
\tablecaption{\ubvri Photometry of SN~2004S\tablenotemark{a}\label{opt_photom}}
\tablehead{   \colhead{JD$-$2,450,000} &
\colhead{$U$} & \colhead {$B$} & \colhead{$V$} &
\colhead{$R$} & \colhead{$I$} & \colhead {Telescope\tablenotemark{b}} }
\startdata
3042.72*       &  14.645 (0.053) & 14.554 (0.024) & 14.447 (0.024) & 14.385 (0.022) & 14.762 (0.021) & 1 \\ 
3045.75*       &  14.965 (0.060) & 14.723 (0.023) & 14.491 (0.019) & 14.446 (0.025) & 14.839 (0.022) & 1 \\ 
3048.76\phm{ } &  15.317 (0.085) & 14.948 (0.029) & 14.592 (0.034) & 14.596 (0.015) & 15.015 (0.024) & 1 \\ 
3052.72*       &  15.865 (0.061) & 15.335 (0.026) & 14.805 (0.024) & 14.823 (0.023) & 15.246 (0.023) & 1 \\ 
3056.68*       &  16.452 (0.051) & 15.799 (0.025) & 15.050 (0.022) & 14.962 (0.021) & 15.228 (0.023) & 1 \\ 

3060.59\phm{1} &  16.886 (0.054) & 16.231 (0.025) & 15.256 (0.022) & 15.040 (0.032) & 15.128 (0.022) & 1 \\ 
3066.67*       &  17.406 (0.061) & 16.744 (0.026) & 15.546 (0.028) & 15.161 (0.026) & 14.993 (0.022) & 1 \\ 
3070.60*       &  17.729 (0.080) & 17.093 (0.031) & 15.808 (0.022) & 15.415 (0.022) & 15.125 (0.028) & 1 \\ 
3074.57*       &  17.881 (0.066) & 17.357 (0.026) & 16.103 (0.025) & 15.723 (0.023) & 15.394 (0.022) & 1 \\ 
3081.64*       &  18.094 (0.063) & 17.600 (0.023) & 16.377 (0.033) & 16.035 (0.022) & 15.750 (0.023) & 1 \\ 

3087.60\phm{ } &  18.330 (0.087) & 17.806 (0.019) & 16.602 (0.015) & 16.306 (0.011) & 16.105 (0.019) & 1 \\ 
3090.55*       &  18.445 (0.047) & 17.841 (0.023) & 16.693 (0.022) & 16.402 (0.021) & 16.247 (0.020) & 1 \\ 
3092.94\phm{ } &  18.301 (0.104) & 17.858 (0.046) & 16.806 (0.037) & 16.473 (0.067) & 16.333 (0.053) & 2 \\

3096.54\phm{ } &  18.528 (0.104) & 17.901 (0.052) & 16.856 (0.050) & 16.614 (0.028) & 16.452 (0.025) & 1 \\ 
3102.55*       &  18.581 (0.147) & 17.964 (0.039) & 16.967 (0.027) & 16.709 (0.027) & 16.711 (0.034) & 1 \\ 
3109.51\phm{ } &  18.786 (0.066) & 18.101 (0.026) & 17.177 (0.020) & 16.967 (0.021) & 16.954 (0.061) & 1 \\ 

3116.50\phm{ } &  18.834 (0.061) & 18.162 (0.022) & 17.325 (0.019) & 17.190 (0.014) & 17.203 (0.052) & 1 \\ 
3122.48*       &  \nodata        & 18.368 (0.066) & 17.573 (0.032) & 17.407 (0.044) & 17.299 (0.045) & 1 \\ 
3130.51*       &  18.790 (0.321) & 18.259 (0.067) & 17.685 (0.033) & 17.638 (0.044) & 17.667 (0.053) & 1 \\ 
3137.47\phm{ } &  19.014 (0.228) & 18.438 (0.045) & 17.811 (0.022) & 17.799 (0.018) & 17.894 (0.040) & 1 \\ 
3151.47\phm{ } &  \nodata        & 18.619 (0.038) & 18.122 (0.034) & 18.214 (0.098) & 18.194 (0.036) & 1 \\ 

3238.31\phm{ } &  \nodata        &  \nodata       & 19.778 (0.058) & 20.369 (0.090) & 19.489 (0.089) & 2 \\
3267.26\phm{ } &  \nodata        &  \nodata       & 20.394 (0.049) & 21.034 (0.080) & 20.547 (0.151) & 2 \\
3357.75\phm{ } &  \nodata        & 21.731 (0.035) & 21.471 (0.041) & 22.271 (0.123) & 21.306 (0.102) & 3 \\
3361.14\phm{ } &  \nodata        &  \nodata       & 21.689 (0.094) & 22.292 (0.247) & 21.217 (0.126) & 2 \\
3453.51\phm{ } &  \nodata        & 23.198 (0.140) & 23.084 (0.117) &  \nodata       &  \nodata       & 3 \\

\enddata
\tablenotetext{a} {The nights marked by an asterisk were photometric at CTIO.  On these occasions the
SN~2004S photometry was tied directly to the \citet{Lan92} system using the stars in Ru 149.  
The numbers in parentheses are 1-$\sigma$ errors (mean errors of the mean)
and include the RMS error of the photometric calibration
from the standards and the scatter of the multiple measurements of
SN~2004S with respect to their mean.  }
\tablenotetext{b}{1 = CTIO 1.3-m.  2 = Siding Spring 2.3-m. 3 = La Silla 2.2-m.}
\end{deluxetable}

\begin{deluxetable}{cccccc}
%\tabletypesize{\scriptsize}
\tablewidth{0pc}
\tablecaption{Near Infrared Photometry of SN~2004S\tablenotemark{a}\label{ir_photom}}
\tablehead{   \colhead{JD$-$2,450,000} & \colhead{$Y$} &
\colhead {$J$} & \colhead{$H$} & \colhead{$K$} & \colhead{Telescope\tablenotemark{b}} }
\startdata
3042.72 &  15.170 (0.026) & 14.945 (0.021) & 15.064 (0.022) & 14.874 (0.059) & 1 \\
3045.75 &  15.340 (0.026) & 15.238 (0.024) & 15.090 (0.021) & 14.976 (0.060) & 1 \\
3048.76 &  15.472 (0.026) & 15.641 (0.029) & 15.228 (0.026) & 15.116 (0.066) & 1 \\
3052.72 &  15.499 (0.026) & 16.156 (0.036) & 15.160 (0.021) & 15.042 (0.063) & 1 \\
3053.43 &    \nodata      & 16.124 (0.010) & 15.151 (0.012) & 14.946 (0.019) & 2 \\
3056.68 &  15.299 (0.025) & 16.258 (0.040) & 15.107 (0.021) & 14.886 (0.048) & 1 \\
3060.59 &  15.181 (0.023) & 16.227 (0.040) & 15.041 (0.021) & 14.957 (0.054) & 1 \\
3070.60 &  14.750 (0.020) & 15.866 (0.036) & 15.050 (0.024) & 15.007 (0.055) & 1 \\
3074.57 &  14.782 (0.020) & 15.932 (0.035) & 15.260 (0.027) & 15.204 (0.060) & 1 \\
3087.60 &  15.544 (0.037) & 17.013 (0.093) & 16.017 (0.046) & 16.017 (0.148) & 1 \\
3090.55 &  15.727 (0.031) & 17.233 (0.099) & 16.221 (0.056) & 16.065 (0.153) & 1 \\
3096.54 &  16.059 (0.055) &    \nodata     &    \nodata     &   \nodata      & 1 \\
3102.55 &  16.293 (0.043) & 18.200 (0.267) & 16.779 (0.085) & 16.561 (0.254) & 1 \\
3109.51 &      \nodata    & 18.587 (0.259) & 16.851 (0.077) &    \nodata     & 1 \\
3116.50 &      \nodata    & 18.652 (0.231) & 17.030 (0.099) &    \nodata     & 1 \\
3130.51 &      \nodata    &    \nodata     & 18.407 (0.320) &    \nodata     & 1 \\
3341.81 &      \nodata    &    \nodata     & 20.57 (0.18)   &    \nodata     & 3 \\ 
3342.80 &      \nodata    & 21.35 (0.06)   &    \nodata     &    \nodata     & 3 \\
3414.71 &      \nodata    & 21.27 (0.13)   &    \nodata     &    \nodata     & 3 \\
3415.62 &      \nodata    &   \nodata      & 20.63 (0.18)   &    \nodata     & 3 \\

\enddata
\tablenotetext{a} {The numbers in parentheses are 1-$\sigma$ uncertainties (mean
errors of the mean).}
\tablenotetext{b} {1 = CTIO 1.3-m. 2 = TNG. 3 = VLT number 1.}
\end{deluxetable}

\begin{deluxetable}{ccccccc}
\tablewidth{0pc}
\tablecaption{Spectra of SN~2004S\label{spectra}}
\tablehead{  \colhead{UT Date} & 
\colhead{Telescope\tablenotemark{a}} &
\colhead{JD$-$2,450,000} & \colhead{Phase\tablenotemark{b}} &
\colhead{Range (\AA)} & \colhead{Resolution (\AA)} & 
\colhead{Exptime (sec)} }
\startdata
Feb06.10 & 1 & 3041.60 &  1.7 & 3100-9650  & 17    & 900   \\
Feb11.96 & 2 & 3047.46 &  7.6 & 3230-10400 & 25    & 900   \\
Feb17.00 & 3 & 3052.50 & 12.6 & 4880-7940  & 20    & 2x600 \\
Feb17.85 & 3 & 3053.35 & 13.5 & 3260-9050  & 20    & 2x600 \\
Feb18.45 & 4 & 3053.95 & 14.1 & 3360-10380 & 10-12 & 2x180 \\
Feb22.92 & 3 & 3058.42 & 18.5 & 3190-9050  & 20    & 2x600 \\
Mar11.88 & 3 & 3076.38 & 36.5 & 3200-9040  & 20    & 2x900 \\
Mar17.85 & 3 & 3082.35 & 42.5 & 3210-9060  & 15    & 2x900 \\

%20040206 02.05  ctio   53041.597   1.7  3100-9650   ...  15min
%20040211 23.08  tng    53047.462   7.6  3230-10400  ~25  15min
%20040216 24.00  not    53052.500  12.6  4880-7940  20 2x10min
%20040218 10.75  keck   53053.948  14.2  3360-10380  ...
%20040217 20.58  not    53053.354  13.5  3260-9050  20   2x10min 
%20040222 21.75  not    53058.417  18.5  3190-9050  20   2x10min
%20040311 20.87  not    53076.375  36.5  3200-9040  20   2x15min
%20040317 20.78  not    53082.346  42.5  3210-9060  15   2x15min
\enddata
\tablenotetext{a} {1 = CTIO 1.5-m. 2 = 3.58-m Telescopio Nazionale Galileo.
3 = 2.56-m Nordic Optical Telescope. 4 = Keck 10-m, using LRIS.}
\tablenotetext{b} {Days since T($B_{max}$) = JD 2,453,039.87.}
\end{deluxetable}

\begin{deluxetable}{ccccc}
\tablewidth{0pc}
\tablecaption{Adjusting SN~2001el templates to the light curves of SN~2004S\label{adjust}}
\tablehead{   \colhead{Filter} &
\colhead{$\Delta$mag$_1$\tablenotemark{a}} & 
\colhead{A$_{\lambda}$/A$_V$\tablenotemark{b}} & 
\colhead{A$_{Gal}$\tablenotemark{c}} &
\colhead{$\Delta$mag$_2$\tablenotemark{d}}
}
\startdata
$U$ & 1.725 (0.062) & 1.5288 & 0.413 (0.041) & 1.312 (0.074) \\
$B$ & 1.608 (0.025) & 1.3147 & 0.355 (0.036) & 1.253 (0.044) \\
$V$ & 1.735 (0.022) & 1.0000 & 0.270 (0.027) & 1.465 (0.035) \\
$R$ & 1.781 (0.019) & 0.8299 & 0.224 (0.022) & 1.557 (0.029) \\
$I$ & 1.859 (0.022) & 0.6050 & 0.163 (0.016) & 1.696 (0.027) \\
$J$ & 1.888 (0.033) & 0.2828 & 0.076 (0.008) & 1.812 (0.034) \\
$H$ & 1.919 (0.024) & 0.1785 & 0.048 (0.005) & 1.871 (0.025) \\
$K$ & 1.928 (0.063) & 0.1154 & 0.031 (0.003) & 1.897 (0.063) \\
\enddata
\tablenotetext{a} {Number of magnitudes the SN~2001el templates must
be shifted to produce the best fits to the SN~2004S data.}
\tablenotetext{b} {From column 4 of Table 8 of \citet{Kri_etal06}.}
\tablenotetext{c} {Adjustments for Galactic extinction, equal to
0.270 mag times the coefficients of column 3.  We assumed R$_V$ = 3.1 for
Galactic dust and uncertainties of 10 percent.}
\tablenotetext{d} {Values of column 2 minus those in column 4.  These
are the equivalent light curve template shifts after correcting for
the effects of Galactic extinction.}
\end{deluxetable}

\begin{deluxetable}{lccc}
\tablewidth{0pc}
\tablecaption{Extinction solutions\label{extinction}\tablenotemark{a}}
\tablehead{   \colhead{Filters} &
\colhead{R$_V$} & 
\colhead{A$_V$\tablenotemark{b}} & 
\colhead{$\Delta \mu$\tablenotemark{c}}
}
\startdata
$BVRI$    & 2.51 (+0.50,$-$0.36) & 0.534 (+0.075,$-$0.057) & 1.994 \\
$BVRIJ$   & 2.02 (+0.30,$-$0.28) & 0.451 (+0.041,$-$0.043) & 1.916 \\
$BVRIJH$  & 2.14 (+0.24,$-$0.22) & 0.470 (+0.027,$-$0.026) & 1.934 \\
$BVRIJHK$ & 2.15 (+0.24,$-$0.22) & 0.472 (+0.025,$-$0.025) & 1.936 \\
\enddata
\tablenotetext{a} {Numbers in parentheses are upper and lower
1-$\sigma$ errors.}
\tablenotetext{b} {Difference of $V$-band extinction of
SNe 2001el and 2004S, in magnitudes.}
\tablenotetext{c} {Implied difference of distance moduli of 
SNe 2001el and 2004S, in magnitudes.}
\end{deluxetable}

\clearpage

\figcaption[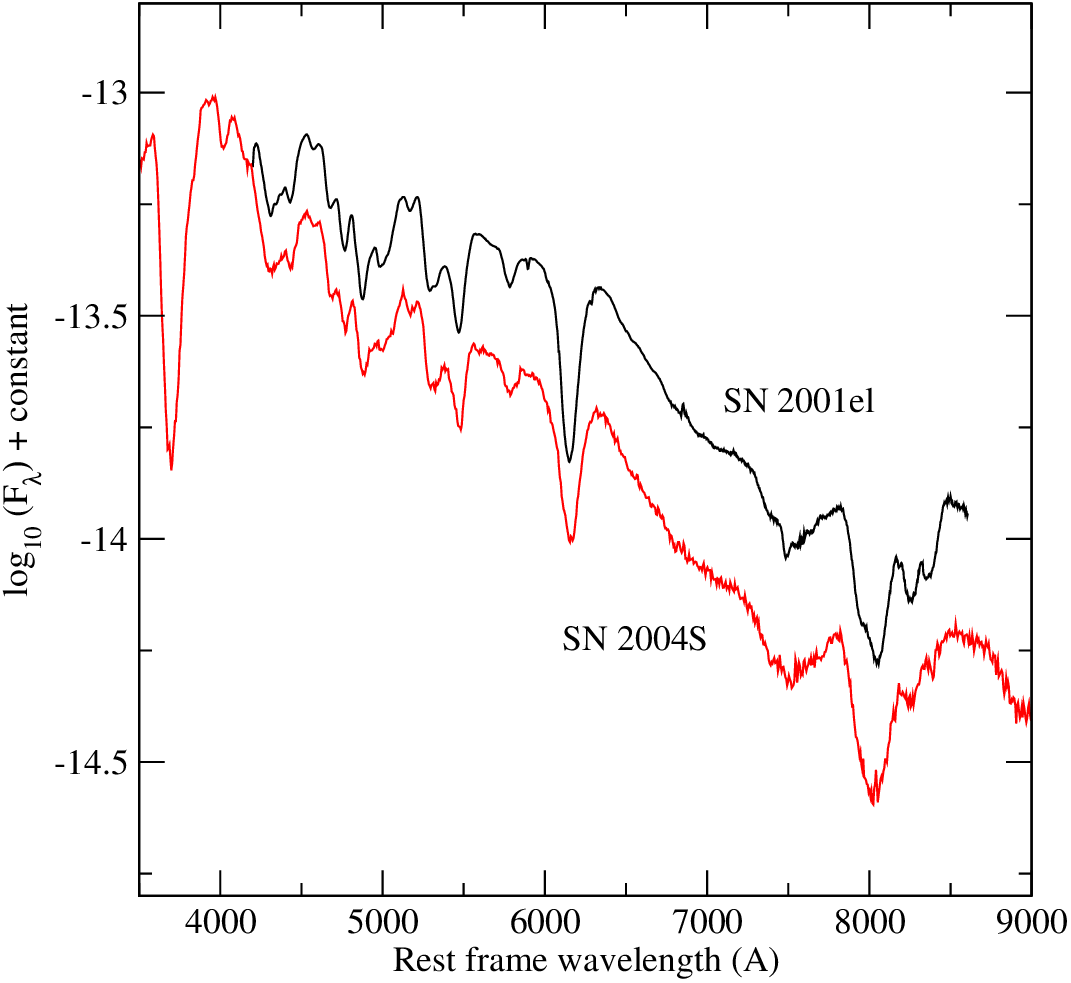] {Maximum light spectra of SN~2001el (2001 October 1), obtained
with the VLT + FORS1, and SN~2004S (2004 February 6), obtained with the CTIO 1.5-m.
The SN~2004S spectrum is shifted by an artibrary amount for display purposes.
\label{comp_at_max}
}

\figcaption[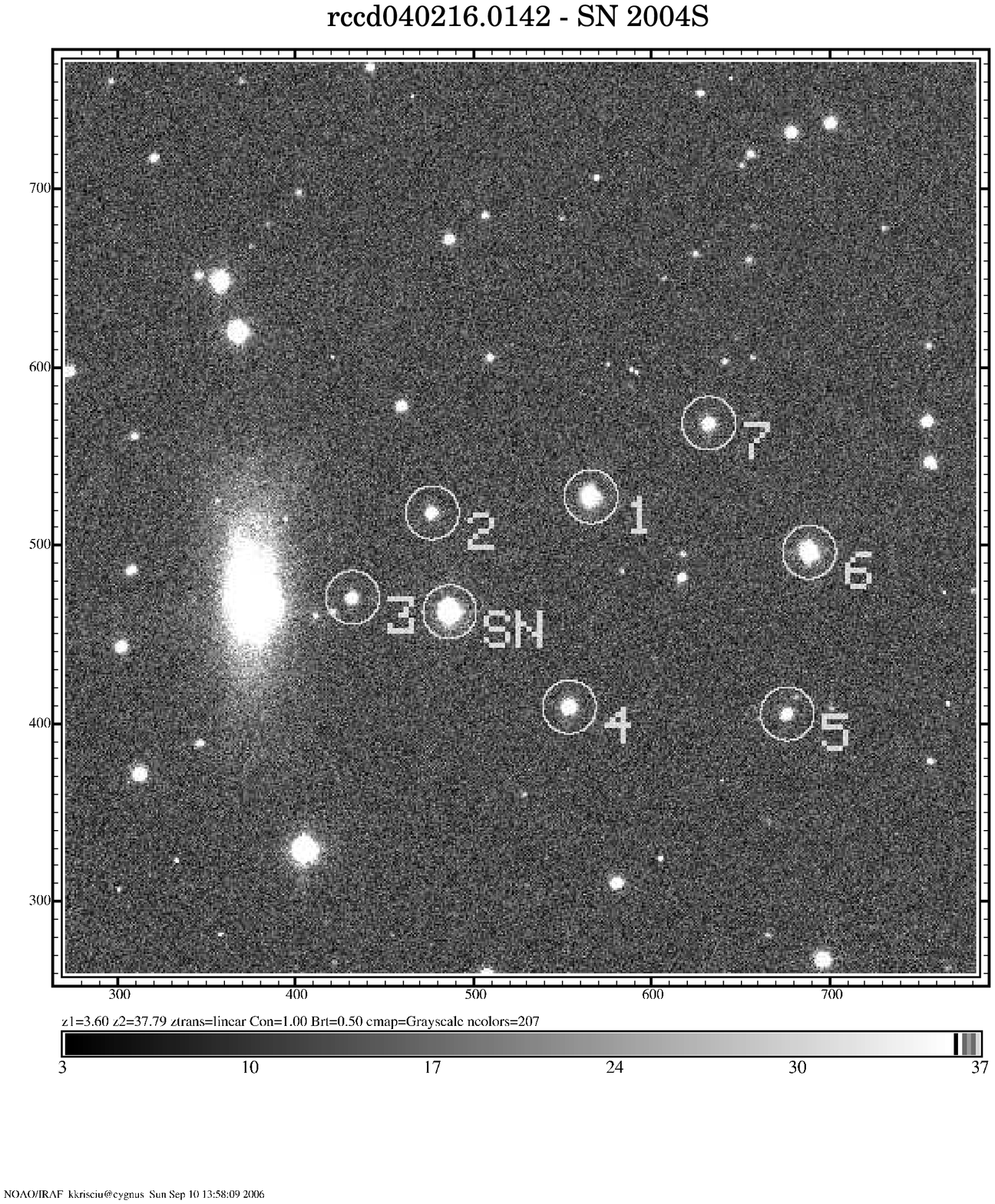]
{MCG -05-16-021, SN~2004S, and the field stars nearby.  This 
3$\farcm$1 by 3$\farcm$1 finder was made from a $V$-band exposure 
obtained with the CTIO 1.3-m telescope on 2004 February 17 UT.
(Note: a different figure will appear in the AJ version of this paper.)
\label{finder}
}

\figcaption[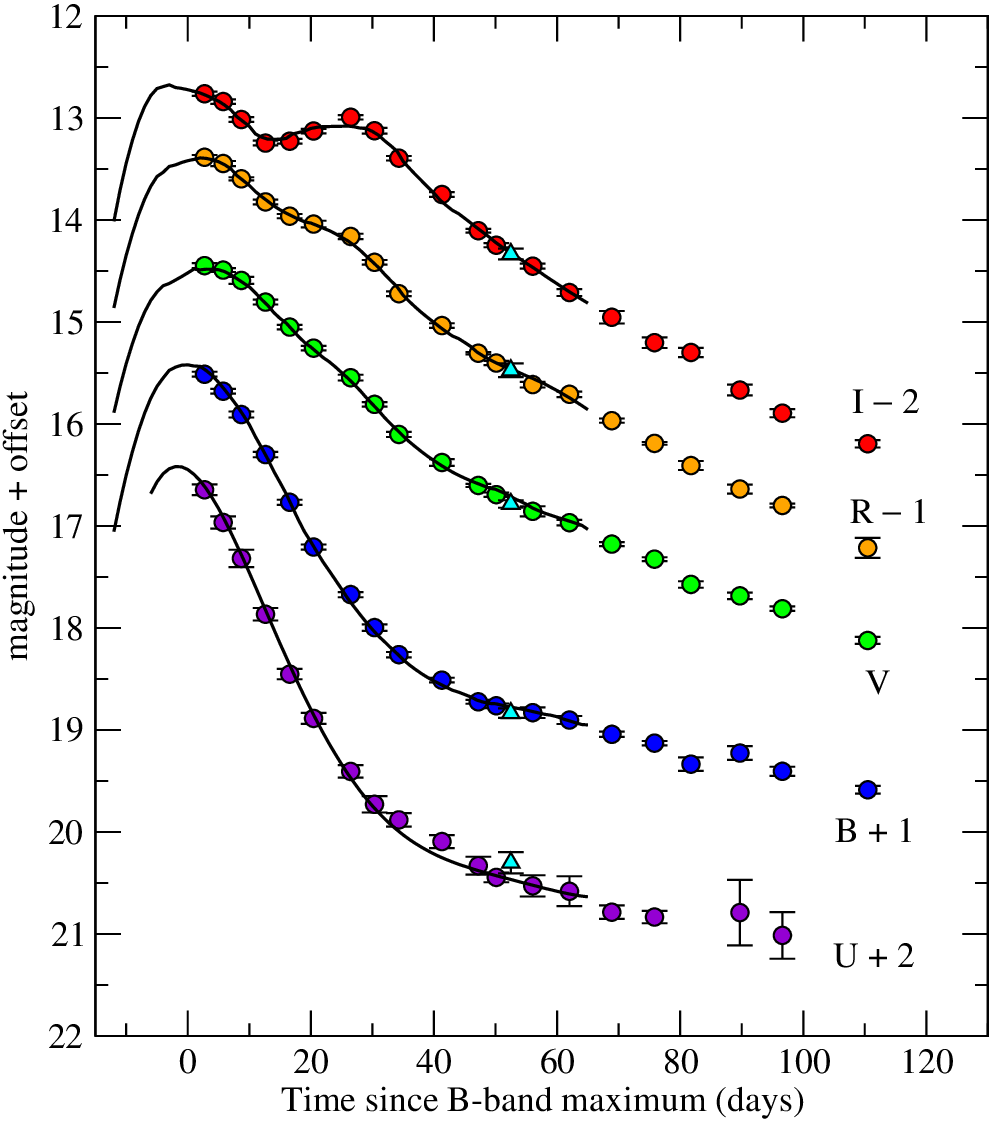] {\ubvri photometry of SN~2004S from 3 to 110
days after T($B_{max}$).  
The circles are data obtained with the CTIO 1.3-m telescope, while the triangles
are data obtained with the Siding Spring 2.3-m telescope.  The templates
are based on the light curves of SN~2001el and have been adjusted in the 
Y-direction to minimize the $\chi^2$ statistic of the fits to the data of 
SN~2004S obtained with the CTIO 1.3-m.  \label{ubvri}
}

\figcaption[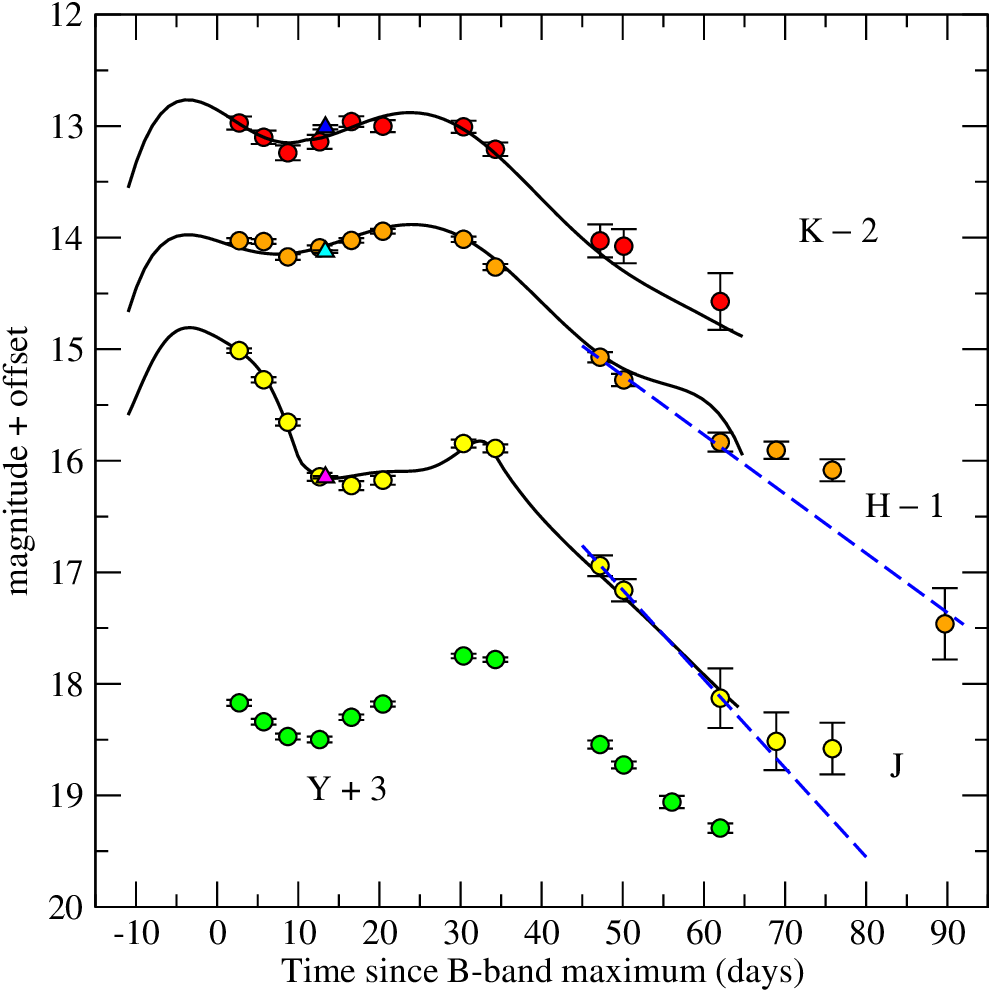] {\yjhk photometry of SN~2004S to 90 days after 
maximum light.
The circles are data obtained with the CTIO 1.3-m telescope, while the triangles
are data from one night with the 3.58-m TNG.  The solid black lines are fits to the
data of SN~2001el, adjusted in the Y-direction to minimize the $\chi^2$ statistic
of the fits to the CTIO 1.3-m data.  The dashed blue lines are estimates of a
linear decline of the light curves.  At $t$ = 69 and 76 days the $H$-band data are
4.3$\sigma$ and 5.3$\sigma$ above the dashed line.  At $t$ = 76 days the $J$-band
datum is 2.8$\sigma$ above the dashed line.  This late time photometry is evidence
that SN~2004S exhibited a third hump in its light curve, in agreement with the
prediction of \citet{Kas06}.
\label{yjhk}
}

\figcaption[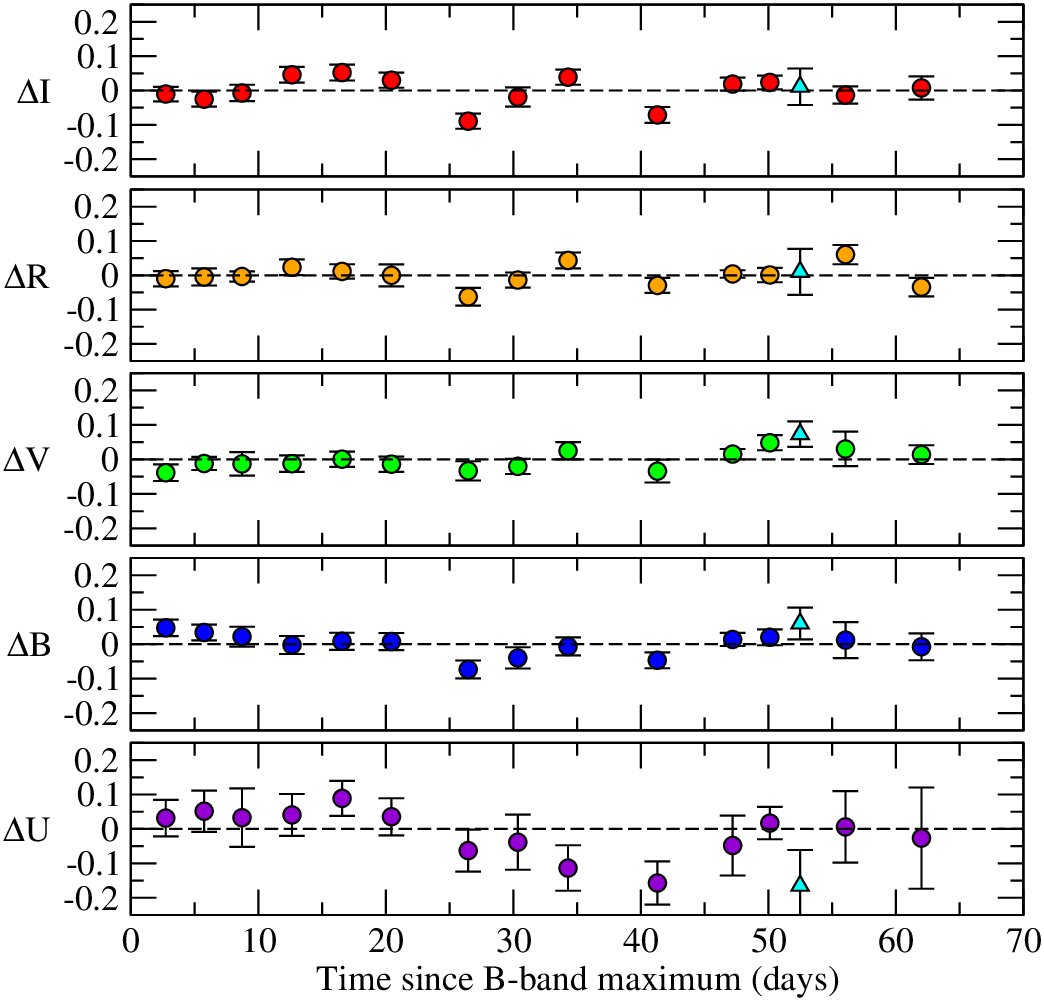] {Residuals of the \ubvri data of
SN~2004S and the light curve templates based on fits to the data of SN~2001el.
``$\Delta$'' is in the sense ``data minus templates''.
\label{opt_res}
}

\figcaption[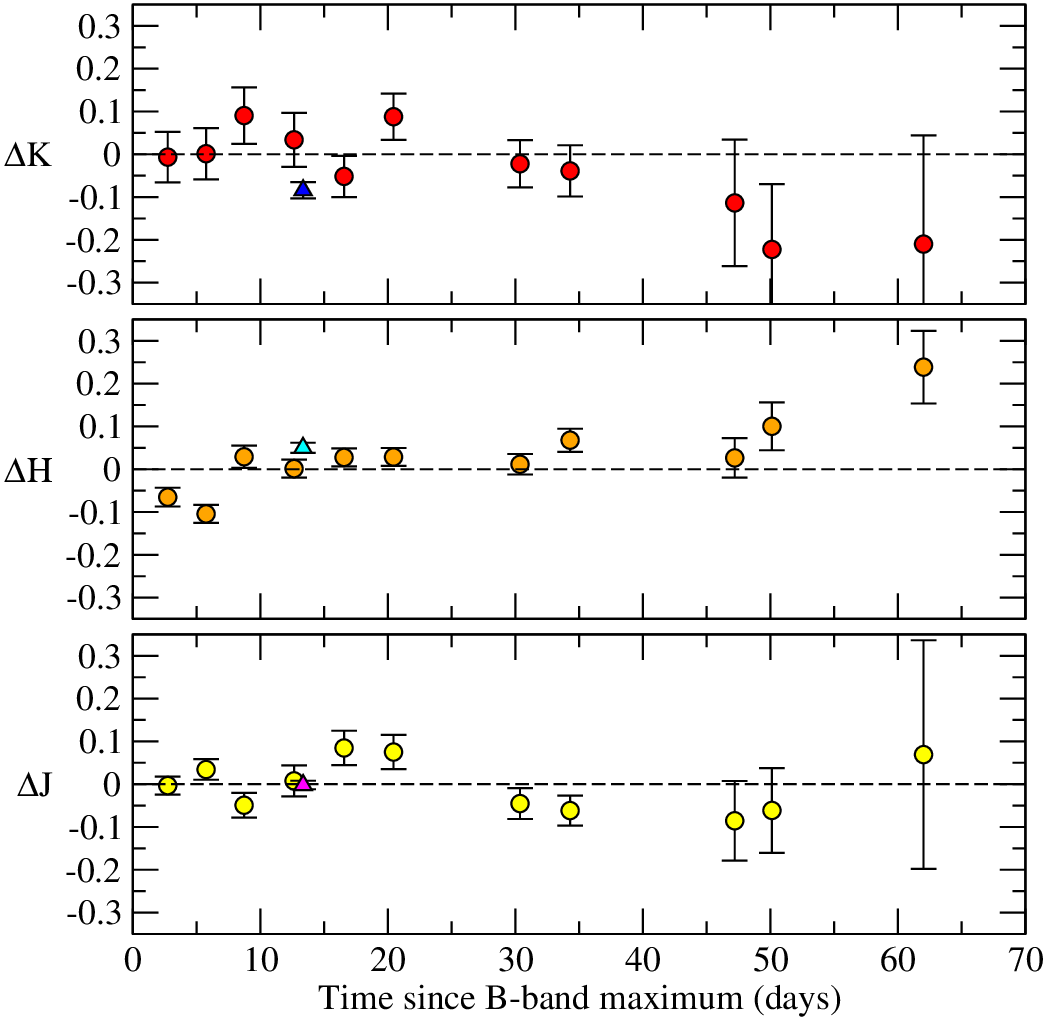] {Residuals of the $JHK$ data of
SN~2004S and the light curve templates based on fits to the data of SN~2001el.
``$\Delta$'' is in the sense ``data minus templates''.
\label{ir_res}
}

\figcaption[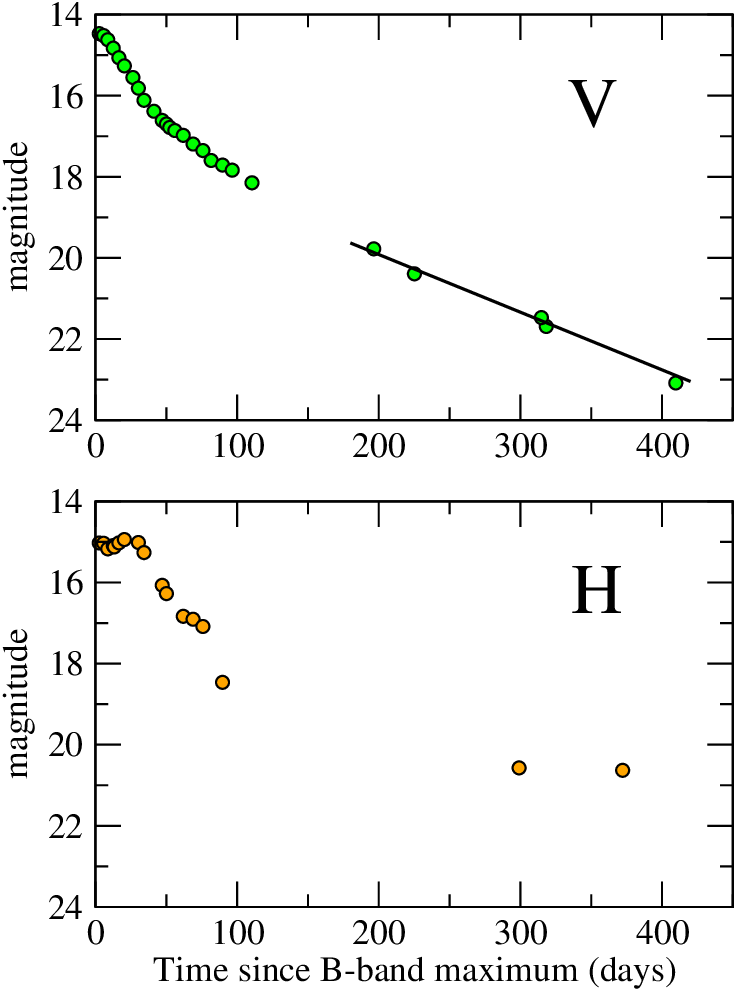] {All of our $V$- and $H$-band of SN~2004S from
Tables \ref{opt_photom} and \ref{ir_photom}.  The photometric errors are 
smaller than or equal to the size of the points.
\label{vh_all}
}

\figcaption[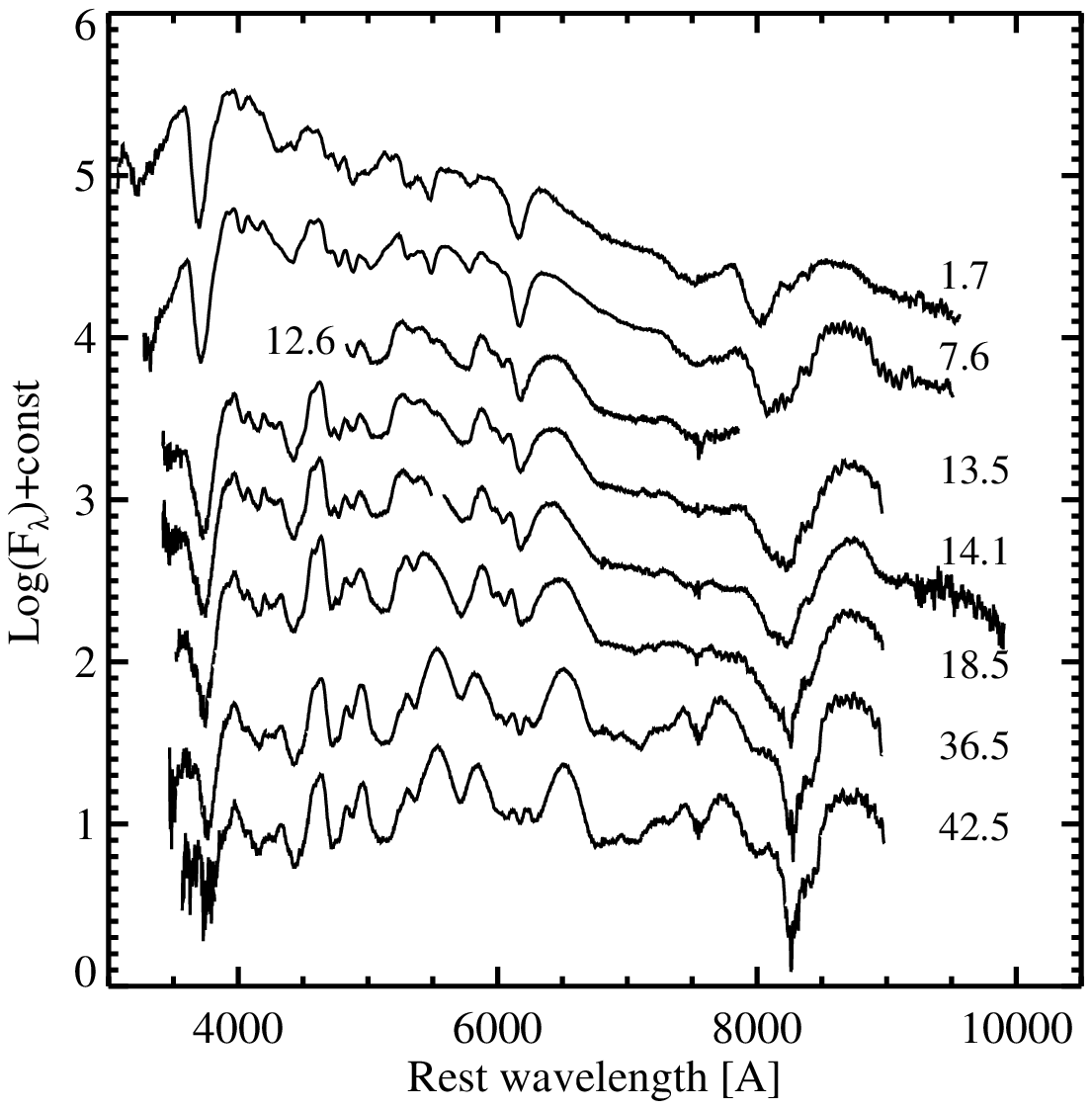] {Optical spectra of SN~2004S from 1.7 to 
42.5 rest frame days after the time of maximum light.  The number of days
since $B$-band maximum is given at the right hand side.  \label{04S_stack}
}

\figcaption[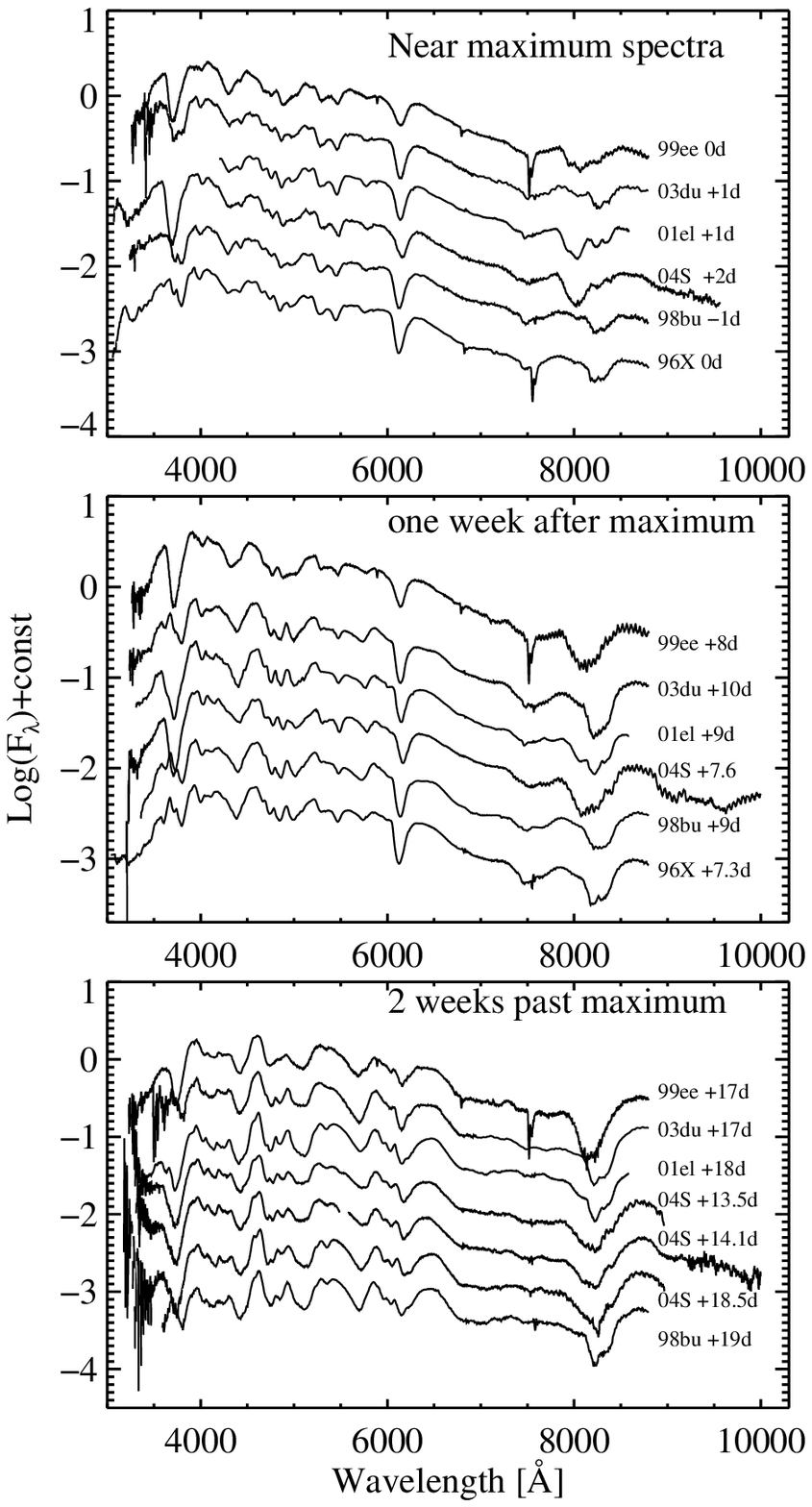] {Comparison of optical spectra of 
various Type Ia SNe and SN~2004S. \label{spectral_comp}
}

\figcaption[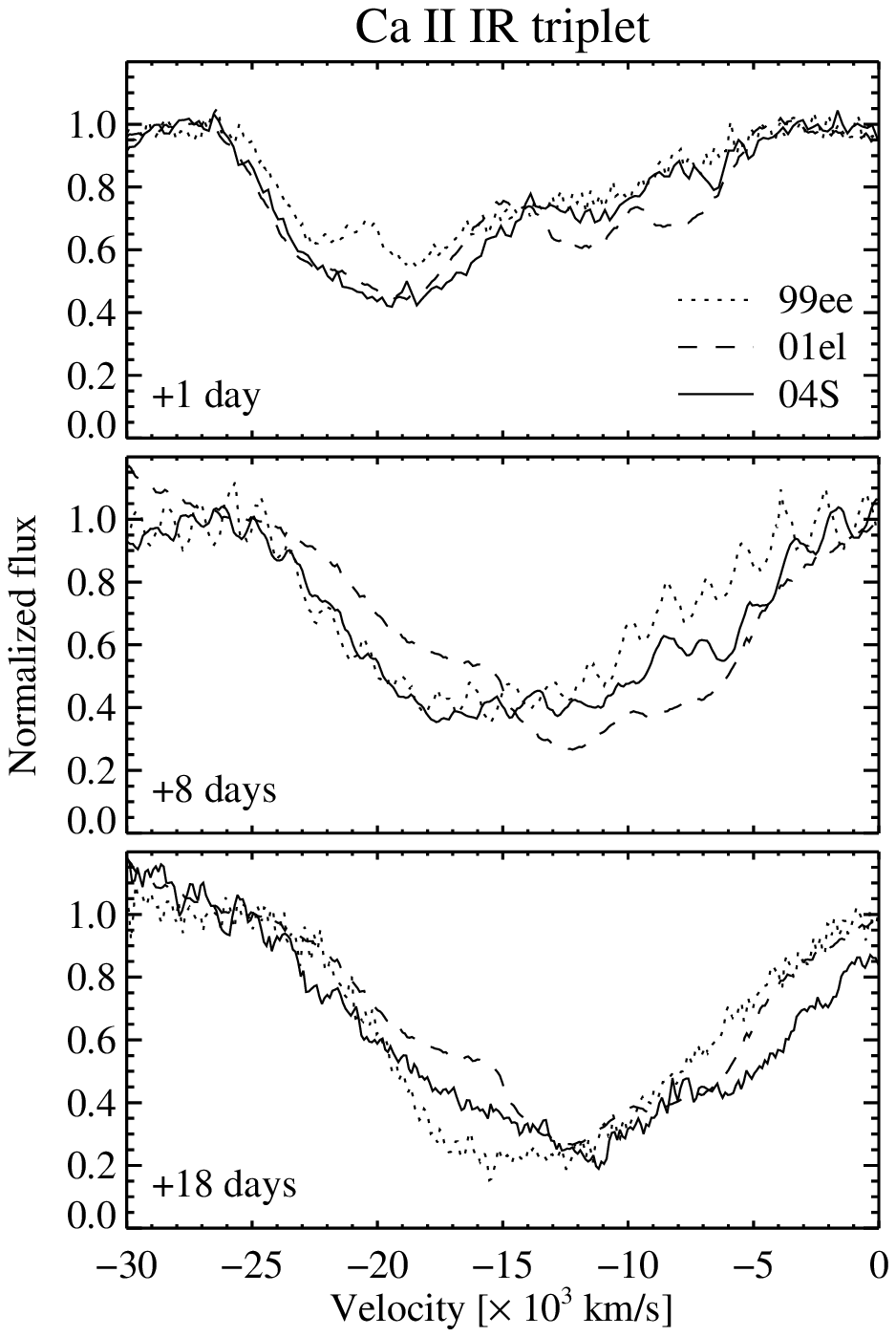] {Evolution of the high-velocity blue shifted
Ca absorption from 1 to 18 days after the time of maximum light of SNe 
1999ee, 2001el, and 2004S.
\label{vel_comp}
}

\figcaption[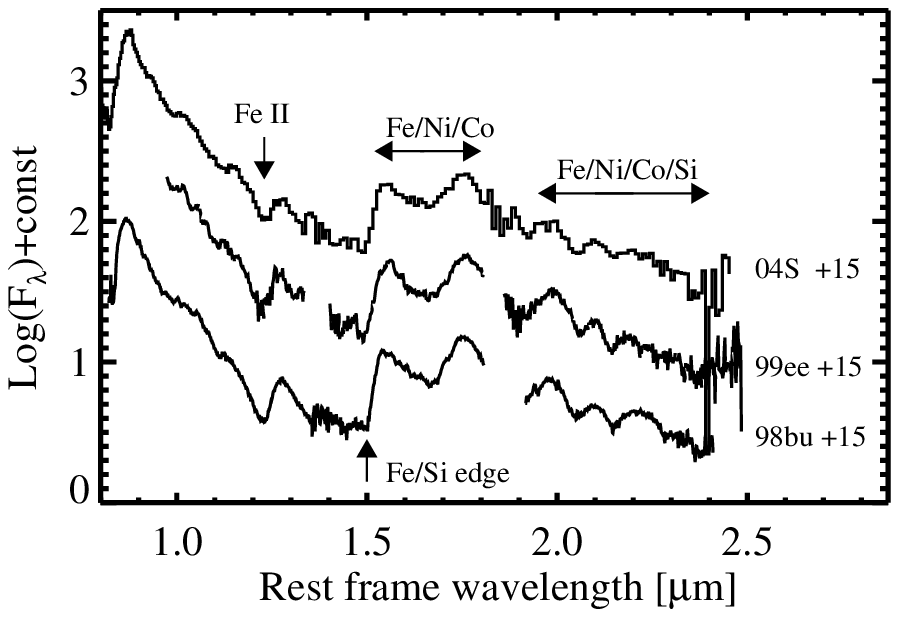] {Comparison of near-IR spectra of the Type Ia SNe
2004S, 1999ee, and 1998bu at 15 days after maximum light. \label{ir_comp}
}

\figcaption[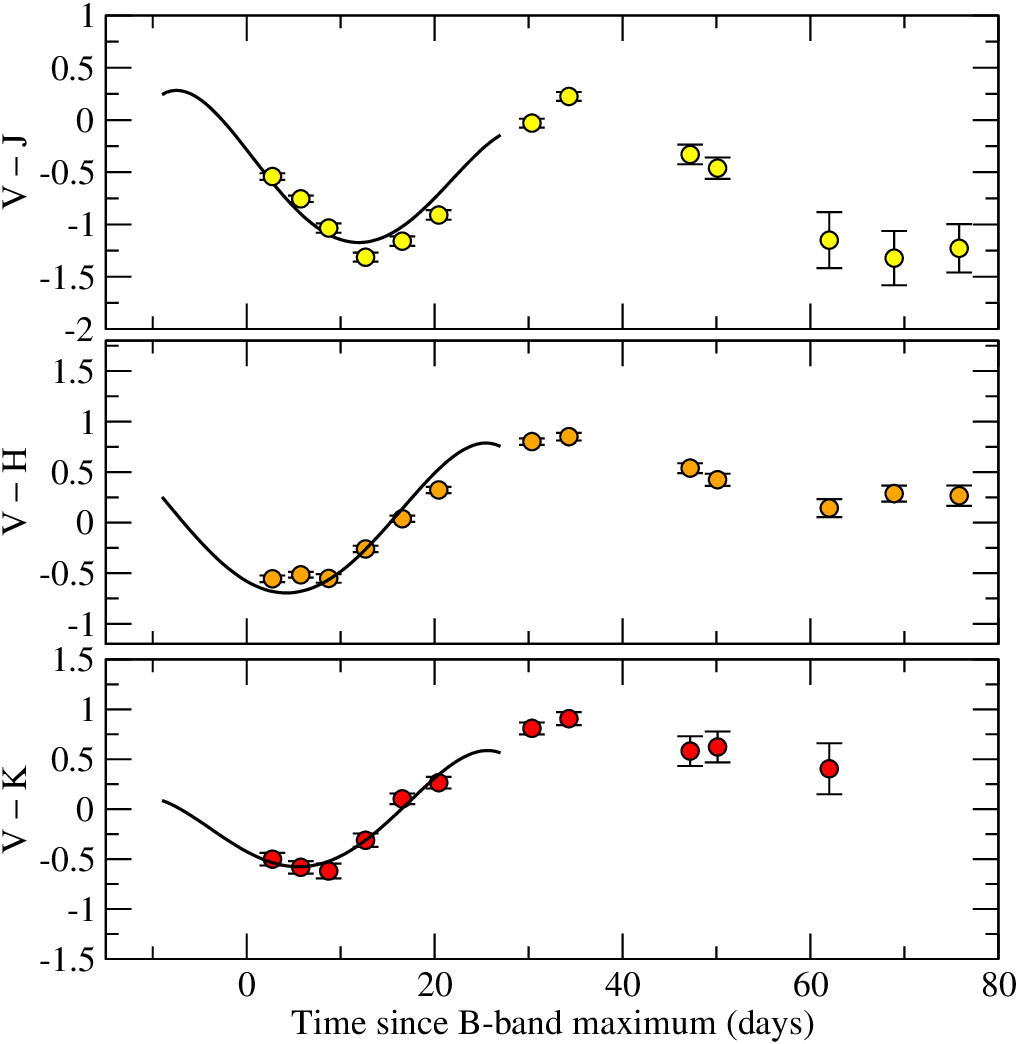] {$V-J$, $V-H$, and $V-K$ color curves of
SN~2004S.  The solid lines are the zero-reddening loci based on eight
Type Ia SNe of \citet{Kri_etal00}, adjusted by the corresponding color
excesses to minimize $\chi^2$.
\label{vir_temp8} 
}

\figcaption[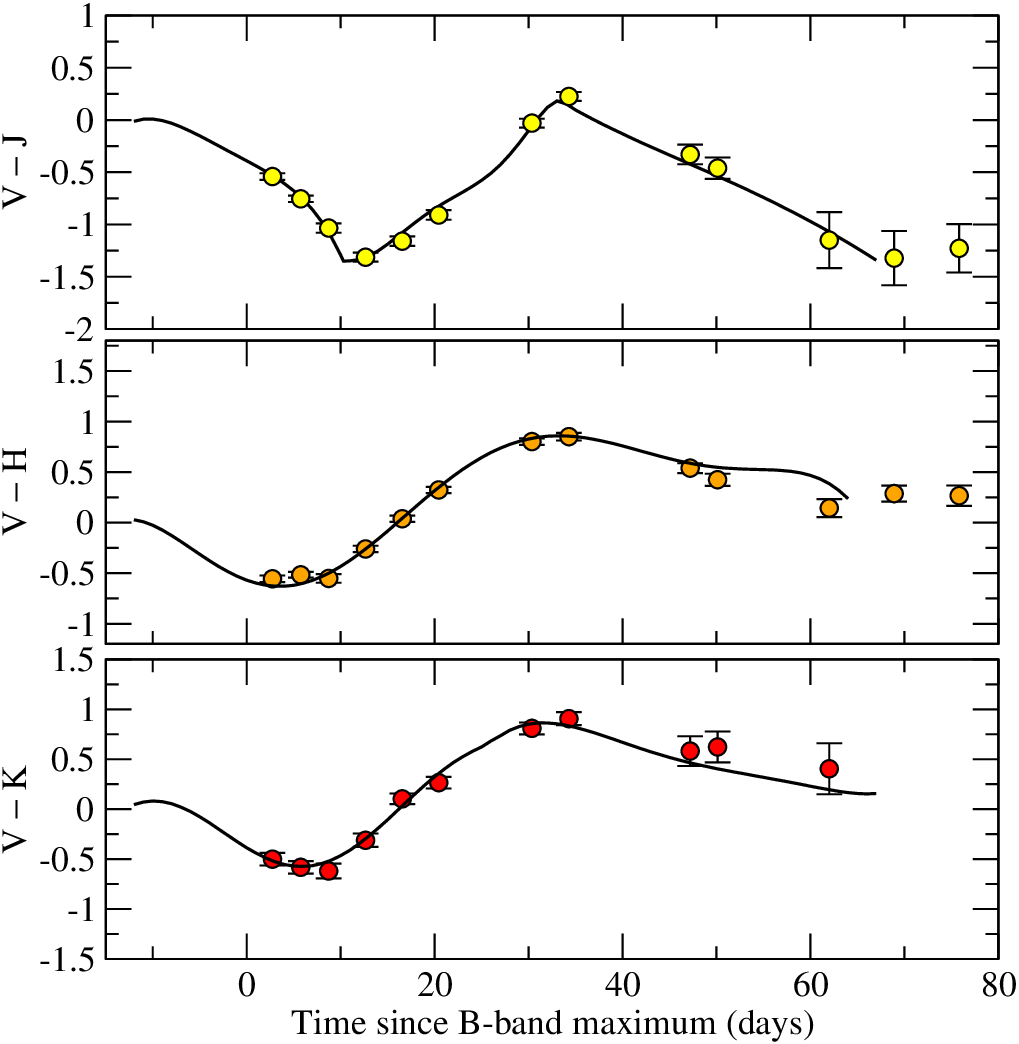] {$V-J$, $V-H$, and $V-K$ color curves of
SN~2004S.  The solid lines are fits to the corresponding data of SN~2001el
and adjusted in the Y-direction to minimize the $\chi^2$ statistic.
\label{vir} 
}

\figcaption[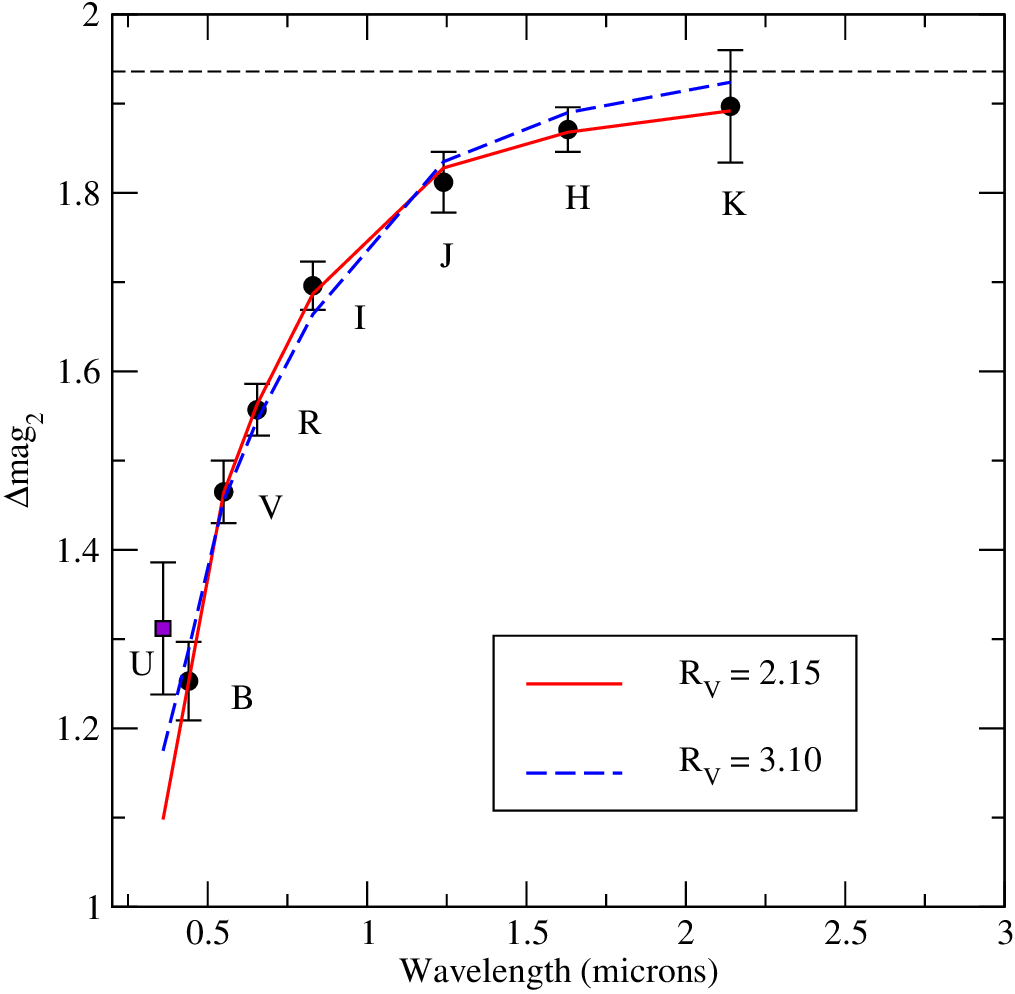] {Values from column 5 of Table \ref{adjust} vs.
the wavelengths of the photometric bands.  In the limit
as the wavelength tends to infinity we obtain a measure of the difference
of the distance moduli of SNe 2001el and 2004S (the horizontal dashed line).
The (red) solid line is the (R$_V$ = 2.15) solution that minimizes the $\chi^2$ 
statistic of the fit.  The (blue) dashed line shows that normal Galactic dust with 
R$_V$ = 3.1 does not fit the data points as well. \label{dmag}
}

\figcaption[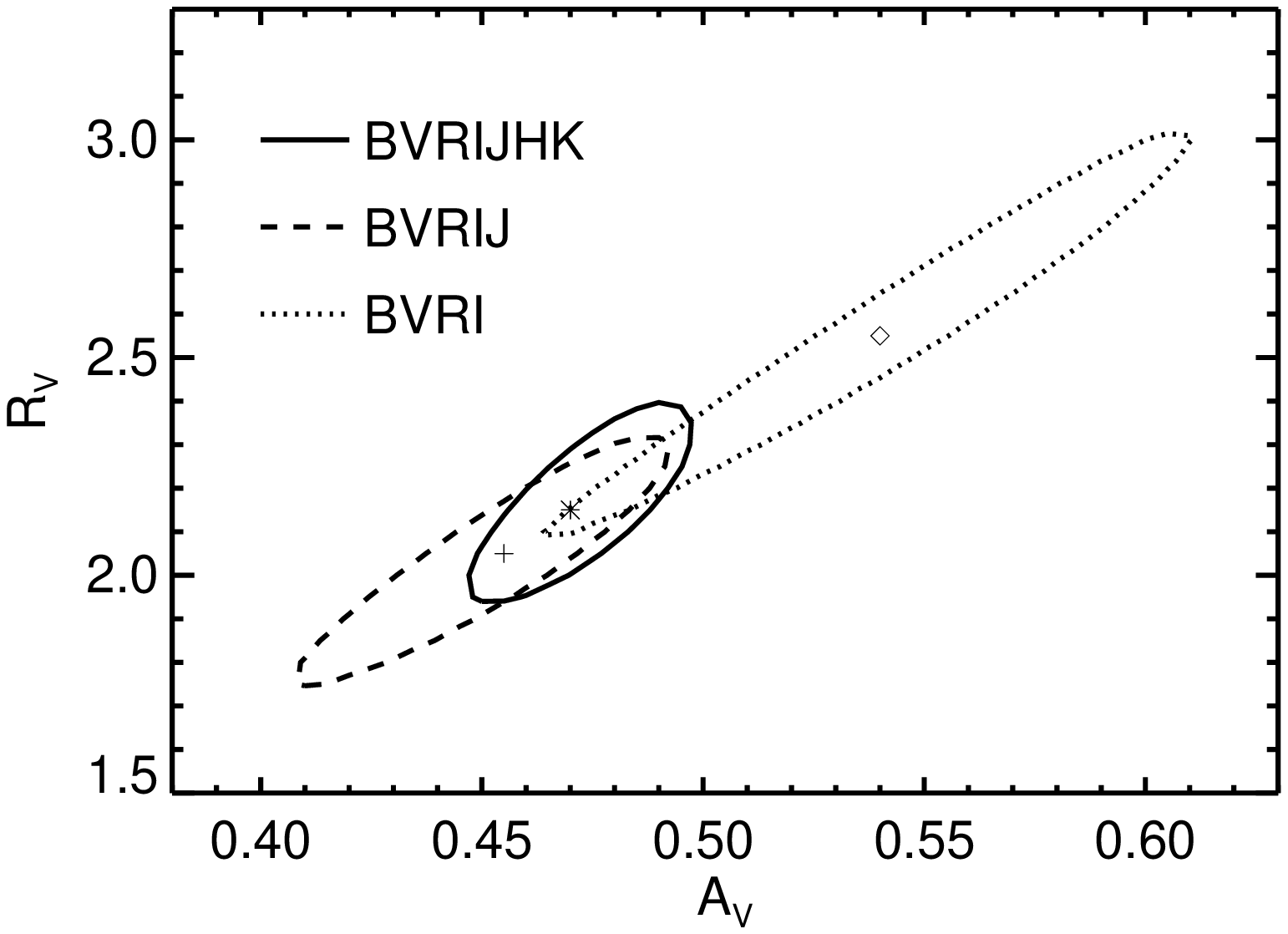] {Contours in the A$_V$-R$_V$ plane.
The X-axis represents possible values of the {\em difference} of
the host galaxy $V$-band extinction of SN~2001el vs. SN~2004S.  The
Y-axis represents the value of R$_V$ pertaining primarily to the 
dust in the host of SN~2001el.  We plot the 68 percent contour levels
for three solutions, using 4, 5, and 7 filter photometry, respectively.
\label{contour}
}

\figcaption[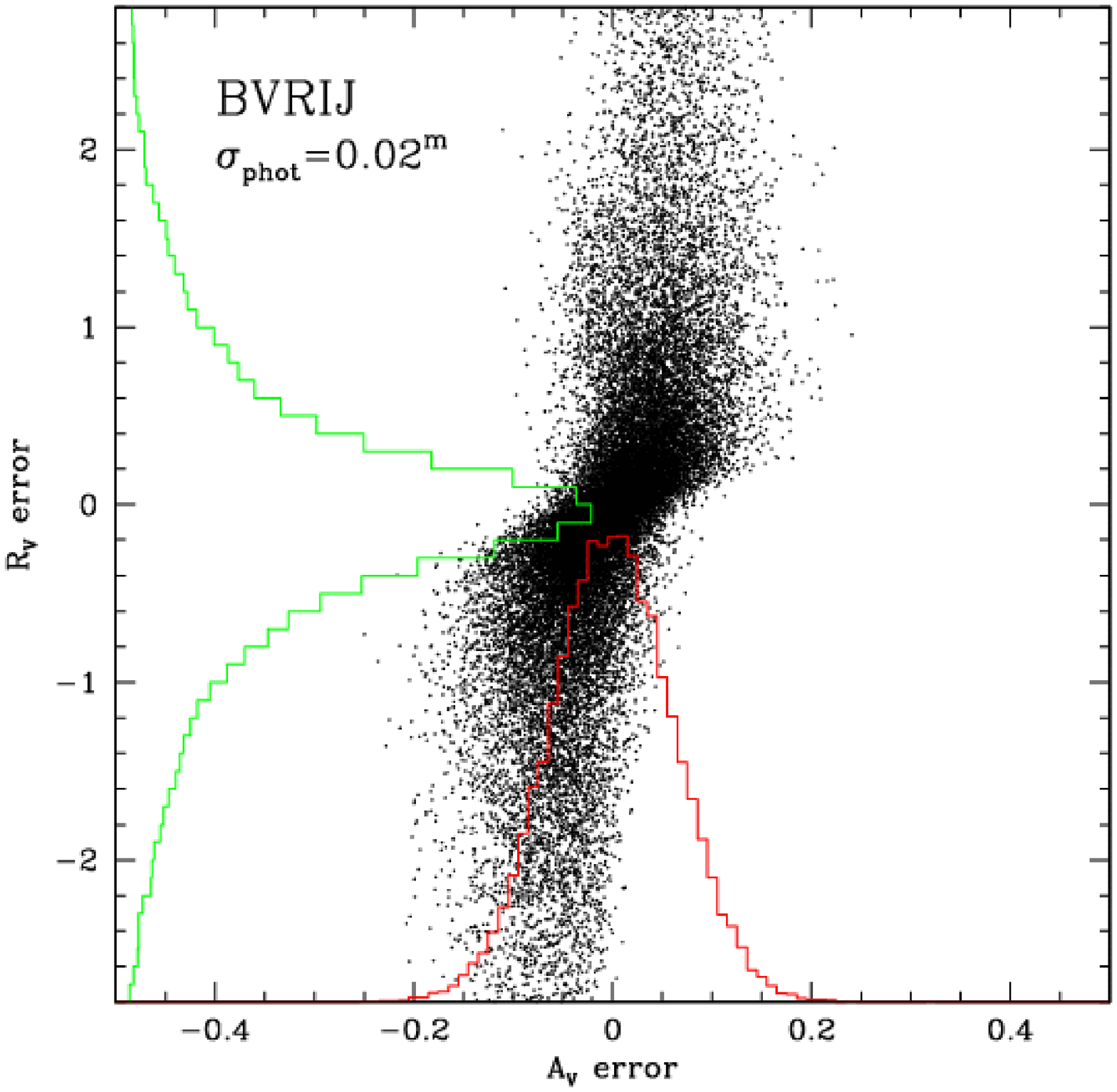] {Correlation of uncertainty in R$_V$ vs.
uncertainty in A$_V$. We have simulated the recovery of the host
galaxy extinction for 30,000 SNe.  For this simulation
we picked at random values of R$_V$ and A$_V$, then recovered them
using \bvrij photometry.  Here we
assumed a $V$-band photometric accuracy of $\pm$ 0.02 mag.
Other bands had correspondingly larger or smaller errors  on the
basis of our actual SN~2001el and SN~2004S data.
\label{scat}
}

\figcaption[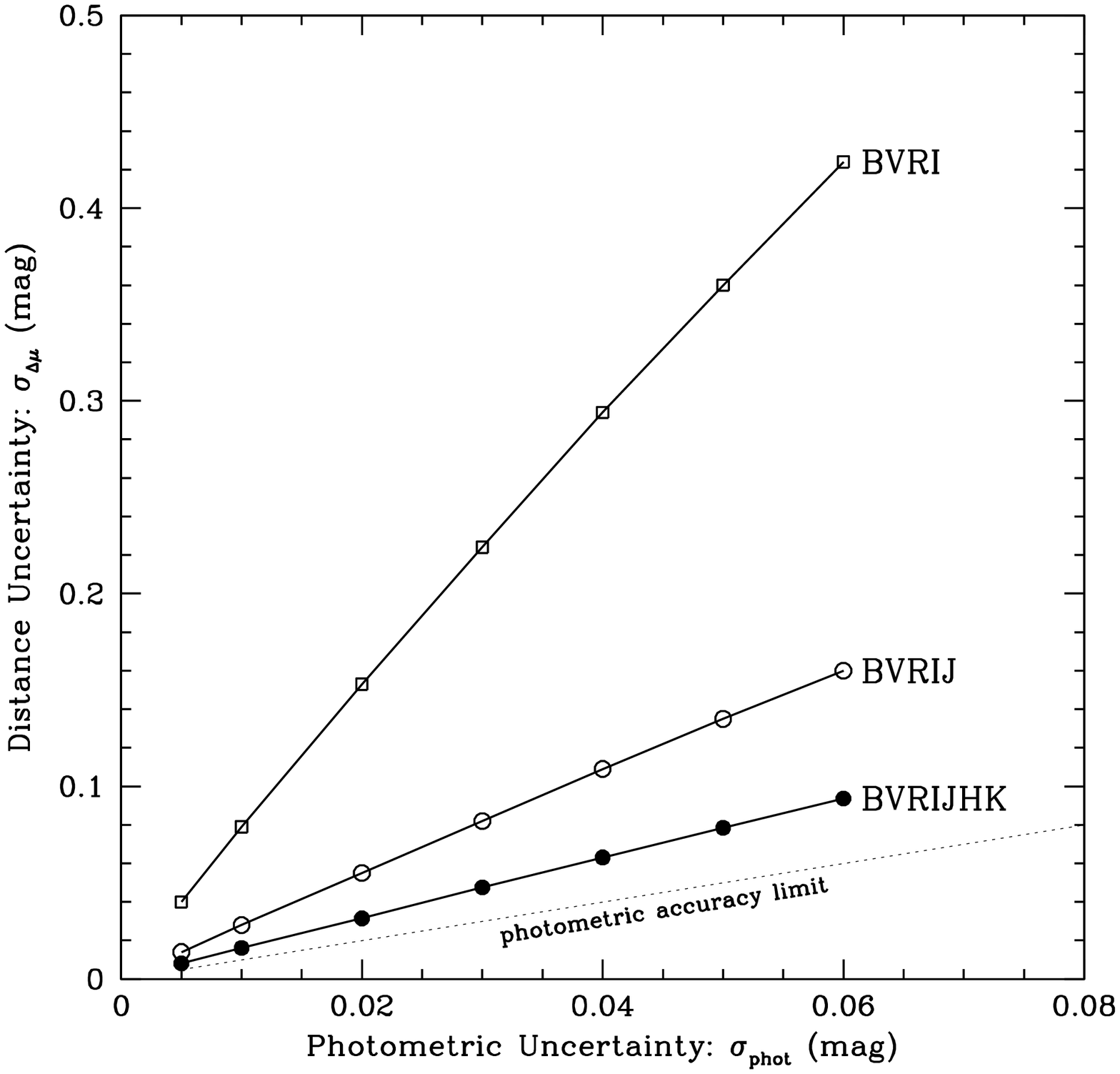] {The uncertainty in the distance modulus for
Type Ia SNe as a function of the photometric accuracy and
the number of filters used to determine the photometric solution.
The dotted line corresponds to the photometric accuracy.  It is
not possible to recover a distance modulus more accurately than
the level of accuracy of the photometry.
\label{mu}
}

\clearpage

\begin{figure}
\plotone{comp_at_max.eps}
{\center Krisciunas {\it et al.} Fig. \ref{comp_at_max}}
\end{figure}

\begin{figure}
\plotone{sn2004S_finder.ps}
{\center Krisciunas {\it et al.} Fig. \ref{finder}}
\end{figure}

\begin{figure}
\plotone{sn2004S_ubvri.eps}
{\center Krisciunas {\it et al.} Fig. \ref{ubvri}}
\end{figure}

\begin{figure}
\plotone{sn2004S_yjhk.eps}
{\center Krisciunas {\it et al.} Fig. \ref{yjhk}}
\end{figure}

\begin{figure}
\plotone{sn2004S_opt_res.eps}
{\center Krisciunas {\it et al.} Fig. \ref{opt_res}}
\end{figure}

\begin{figure}
\plotone{sn2004S_ir_res.eps}
{\center Krisciunas {\it et al.} Fig. \ref{ir_res}}
\end{figure}

\begin{figure}
\plotone{vh_all.eps}
{\center Krisciunas {\it et al.} Fig. \ref{vh_all}}
\end{figure}

\begin{figure}
\plotone{sn2004s_new.eps}
{\center Krisciunas {\it et al.} Fig. \ref{04S_stack}}
\end{figure}

\begin{figure}
\plotone{sn2004s.comp.eps}
{\center Krisciunas {\it et al.} Fig. \ref{spectral_comp}}
\end{figure}

\begin{figure}
\plotone{sn2004s.comp_vel.eps}
{\center Krisciunas {\it et al.} Fig. \ref{vel_comp}}
\end{figure}

\begin{figure}
\plotone{ir_comp_04s.eps}
{\center Krisciunas {\it et al.} Fig. \ref{ir_comp}}
\end{figure}

\begin{figure}
\plotone{sn2004S_vir_temp8.eps}
{\center Krisciunas {\it et al.} Fig. \ref{vir_temp8}}
\end{figure}

\begin{figure}
\plotone{sn2004S_vir.eps}
{\center Krisciunas {\it et al.} Fig. \ref{vir}}
\end{figure}

\begin{figure}
\plotone{dmag.eps}
{\center Krisciunas {\it et al.} Fig. \ref{dmag}}
\end{figure}

\begin{figure}
\plotone{contour.ps}
{\center Krisciunas {\it et al.} Fig. \ref{contour}}
\end{figure}

\begin{figure}
\plotone{scat.ps}
{\center Krisciunas {\it et al.} Fig. \ref{scat}}
\end{figure}

\begin{figure}
\plotone{mu.ps}
{\center Krisciunas {\it et al.} Fig. \ref{mu}}
\end{figure}

\end{document}